\documentclass[a4paper, 10pt, conference]{ieeeconf}      % Use this line for a4 paper

\IEEEoverridecommandlockouts                              % This command is only needed if 
\usepackage{amsmath} % assumes amsmath package installed
\usepackage{amssymb}  % assumes amsmath package installed

\usepackage[acronym]{glossaries}
\newacronym{uav}{UAV}{Unmanned Aerial Vehicle}
\newacronym{ekf}{EKF}{Extended Kalman Filter}
\newacronym{esc}{ESC}{Electronic Speed Controller}
\newacronym{ukf}{UKF}{Unscented Kalman Filter}
\newacronym{imu}{IMU}{Inertial Measurement System}
\newacronym{gps}{GPS}{Global Positioning System}
\newacronym{indi}{INDI}{Incremental Nonlinear Dynamic Inversion}
\newacronym{tcn}{TCN}{Temporal Convolutional Network}
\newacronym{rpm}{RPM}{Revolutions per Minute}
\newacronym{bem}{BEM}{Blade Element Momentum}
\newacronym{rmse}{RMSE}{Root Mean Square Error}
\newacronym{lasso}{LASSO}{Least Absolute Shrinkage and Selection Operator}
\newacronym{rps}{RPS}{revolutions per second}
\newacronym{ned}{NED}{North-East-Down}
\newacronym{ls}{LS}{Least Squares}
\newacronym{nrmse}{nRMSE}{normalized Root Mean Square Error}
\newacronym{rls}{RLS}{Recursive Least Squares}
\newacronym{tow}{TOW}{Take-off Weight}
\newacronym{gvf}{GVF}{Guiding Vector Field}
\newacronym{vtol}{VTOL}{Vertical Takeoff and Landing}

\usepackage{booktabs}   % toprule midrule etc for tables
\usepackage{makecell}
\usepackage{multirow}

\usepackage{overpic}
\usepackage{tikz}
\usetikzlibrary{arrows.meta}

\usepackage{float}
\usepackage{xcolor}

\usepackage{subcaption}

\usepackage{siunitx}
\DeclareSIUnit{\rpm}{RPM}

\usepackage{algorithm}
\usepackage{algpseudocode}

\newcommand{\norm}[1]{\left\lVert #1 \right\rVert}

\makeatletter
\let\NAT@parse\undefined
\makeatother
\usepackage[hidelinks]{hyperref}
\glsdisablehyper

\usepackage[nameinlink,capitalise]{cleveref}  
\crefname{equation}{}{} % remove "eq." before number

\title{\LARGE \bf
Guiding vector field-based guidance under wind disturbances applied to a tailsitter UAV
}

\author{Evangelos Ntouros$^{*}$ and Ewoud J. J. Smeur$^{}$% <-this % stops a space
\thanks{$^{*}$Correspondence {\tt\small e.ntouros@tudelft.nl}}%
}

\begin{document}

\maketitle
\thispagestyle{empty}
\pagestyle{empty}

%%%%%%%%%%%%%%%%%%%%%%%%%%%%%%%%%%%%%%%%%%%%%%%%%%%%%%%%%%%%%%%%%%%%%%%%%%%%%%%%
\begin{abstract}
This paper develops a guidance control law based on a parametric \gls{gvf} and integrates it with a state-of-the-art acceleration and attitude control architecture for tailsitters. The resulting framework enables a direct comparison between traditional trajectory-tracking guidance and \gls{gvf}-based path-following guidance using a realistic tailsitter model operating under windy conditions. Through extensive simulations, it is shown that for agile flight scenarios with wind and small initial position error, both guidance strategies achieve comparable tracking performance, indicating that the additional complexity introduced by the \gls{gvf} formulation is not always justified. However, the \gls{gvf}-based approach exhibits an advantage when initial deviation from the path is present, yielding smooth and well-behaved convergence toward the desired path. Two additional contributions support this evaluation. First, a modification of the parametric \gls{gvf} is proposed that guarantees exponential stability of the tracking error dynamics for a single integrator system. Second, the differential flatness transform of a tailsitter vehicle is extended to account for explicit knowledge of the wind velocity vector.
\newline

\end{abstract}

\glsresetall

%%%%%%%%%%%%%\textbf{}%%%%%%%%%%%%%%%%%%%%%%%%%%%%%%%%%%%%%%%%%%%%%%%%%%%%%%%%%%%%%%%%%%%
\section{INTRODUCTION}

\gls{vtol} aircraft combine the hover capability of rotorcraft with the higher cruise speed and aerodynamic efficiency of fixed-wing aircraft. A specific subclass of \gls{vtol} platforms is the tailsitter \gls{uav}, which transitions from hover to forward flight by pitching down approximately $90^\circ$. Tailsitters may employ tilting motors to produce pitching moment; however, a mechanically simpler configuration relies on only two fixed motors and two control surfaces acting simultaneously as elevators and ailerons, commonly referred to as elevons. This configuration results in a mechanically simple, lightweight, and cost-effective platform, making tailsitters attractive for a wide range of applications, including search and rescue and package delivery.

Operations are typically conducted in challenging environments which introduce two major challenges. First, obstacles and dynamic conditions demand agile flight capabilities to avoid hazards by accurately tracking high-acceleration maneuvers. Second, wind disturbances pose a significant challenge for this class of \glspl{uav}. Due to their large wing surface and limited control authority provided by the elevons, tailsitters are particularly sensitive to external disturbances, which can affect stability and tracking performance.

For the guidance problem, trajectory-tracking approaches are traditionally employed, where a trajectory-generation algorithm designs a time-parameterized reference to be followed based on specified objectives and constraints. For example, Tal et al. \cite{tal2023} generate trajectories for a tailsitter that connect desired waypoints while accounting for feasibility through its dynamics. The resulting trajectories are obtained via snap minimization \cite{mellinger2011}.

On the other hand, path-following algorithms provide a viable alternative in scenarios where strict time constraints are not present. There, the objective is to converge to and follow a prescribed geometric path independently of a specific time parameterization. Among such approaches, guidance methods based on \glspl{gvf} have attracted significant attention in recent years.

The \gls{gvf} framework constructs a velocity vector field such that the \gls{uav} converges to the desired path when tracking that field. Consequently, the resulting guidance law is a function of the vehicle position rather than time. Feedforward terms can therefore be computed based on the current position instead of a time reference, which can improve robustness in the presence of large deviations from the path.

Researchers in \cite{yuri2018} and \cite{demarina2017} developed guidance control laws based on a \gls{gvf} for fixed-wing aircraft. Subsequently, Yao et al.~\cite{yao2021} proposed a parameterized formulation of the \gls{gvf}, which simplifies the vector field design process to a level comparable to conventional trajectory generation. This formulation also eliminates singularities inherent to classical \gls{gvf} constructions, such as the zero-velocity condition at the center of circular paths. The approach was demonstrated on fixed-wing aircraft using separate horizontal and vertical control loops and operating at constant airspeed. Finally, Zhou et al. in \cite{bautista2025} extended the non-parametric \gls{gvf} from \cite{yuri2018} to achieve, under some assumptions, exponentially converging tracking error dynamics that can be easily tuned as a first order system.

Regarding wind disturbances, Smeur et al. \cite{ewoud2016} showed that employing an \gls{indi} attitude controller on a rotorcraft significantly improves robustness due to the incremental and sensor-based nature of the control algorithm, where unmodelled moments are measured and compensated for in the subsequent control loop. This approach was later extended to tailsitter platforms in~\cite{ewoud2020}. Pfeifle et al. \cite{pfeifle2021} proposed an \gls{indi}-based controller for a fixed-wing aircraft formulated in the aerodynamic frame rather than the body frame, enabling improved wind compensation while following spline-based trajectories.

The aforementioned works primarily focus on rejecting wind-induced disturbances through feedback control, rather than explicitly exploiting wind estimates in a feedforward manner to reduce the corrective effort required from the controller. In contrast, Zou et al.~\cite{zou2025} derived the differential flatness transform of a tailsitter with explicit wind knowledge. However, their approach relies on a classical Buckingham $\pi$-theory formulation \cite{anderson2016} of the wing aerodynamics, which relies on extensive and costly wind-tunnel campaigns to identify the underlying aerodynamic model. Furthermore, this formulation suffers from singularities in near-hover flight conditions.

On the other hand, Tal et al.~\cite{tal2022} employed the $\phi$-theory parameterization~\cite{lustosa2019}, which provides a computationally simpler and singularity-free aerodynamic model while effectively capturing the fundamental physics of the tailsitter in a unified way across the flight envelope. The coefficients of this model can be readily identified using experimental flight data. By combining this model with previously developed control architectures and exploiting the differential flatness properties of the system, they calculated appropriate feedforward signals to reduce the tracking error during aggressive aerobatic maneuvers.

The main contribution of this work is the formulation of a guidance controller based on a \gls{gvf} and its consistent integration with the state-of-the-art acceleration and attitude control architecture developed in~\cite{tal2022}. To the best of the authors’ knowledge, this work is the first to integrate a \gls{gvf}-based guidance law with a full three dimensional acceleration controller, whereas most existing approaches rely on partial or decoupled control schemes that treat horizontal and vertical motion separately~\cite{demarina2017, yao2021}. The resulting performance is assessed under unmodelled wind disturbances and initial path deviations and systematically compared against the traditional trajectory-tracking guidance approach presented in~\cite{tal2022}. Conclusions are then drawn regarding the relative performance of the two methods and the conditions under which each approach is preferable.

All evaluations are conducted in numerical simulation using a $\phi$-theory aerodynamic model of the Cyclone tailsitter \gls{uav} (\cref{photo_cyclone}), identified from real flight data, thereby ensuring realistic flight behavior. It is worth noting that although the platform considered in this work is a tailsitter, the developed guidance law is general and applicable to other \gls{uav} configurations. 

Additionally, this work extends the differential flatness transform for tailsitters introduced in~\cite{tal2022} to explicitly account for estimated wind. This extension enables the exploitation of wind estimates, when available, in a feedforward manner to actively counteract wind disturbances.

The final contribution of this work is a modification to the parametric \gls{gvf} formulation presented in~\cite{yao2021}, which guarantees exponential stability of the tracking error dynamics for a single integrator system. This result is essential for the correct tuning and consistent comparison of the two guidance strategies.

\section{TAILSITTER MODEL}\
\label{sec:model}
\subsection{Reference frames}
\label{subsec:frames}
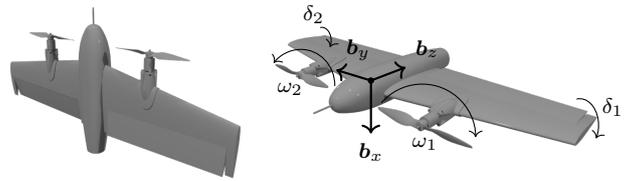
\begin{figure}[t]
    \centering

    \begin{subfigure}[t]{0.41\columnwidth}
        \centering
        \includegraphics[width=\linewidth]{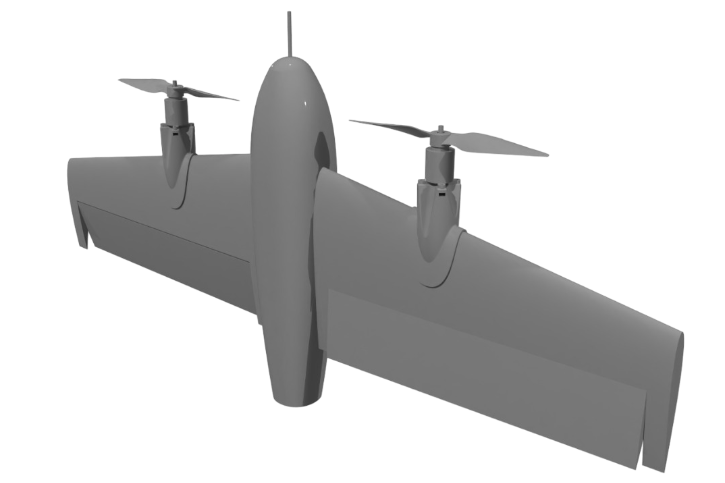}
        \caption{Cyclone during hover.}
        \label{photo_cyclone_hover}
    \end{subfigure}
    \begin{subfigure}[t]{0.515\columnwidth}
        \centering
            \begin{overpic}[width=\columnwidth]{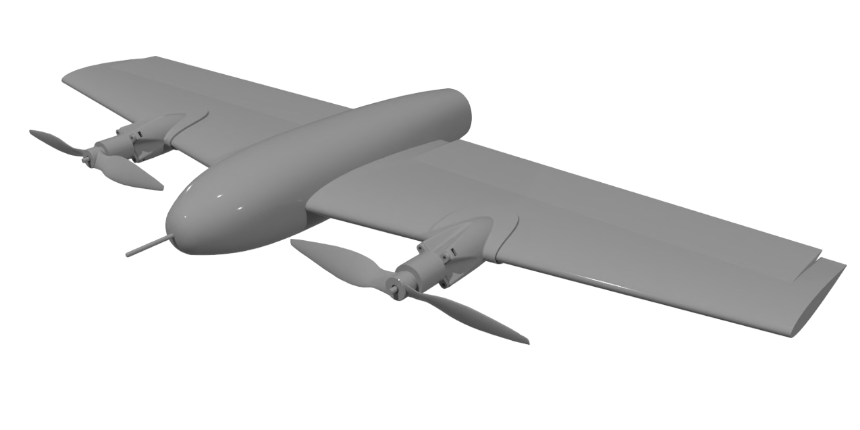}
                \begin{tikzpicture}[overlay, x=1pt, y=1pt]
                
                    % Center of mass
                    \coordinate (O) at (40, 38);
                    \draw[fill=black] (O) circle (1.0pt);
                
                    % Body axes
                    \draw[->, thick] (O) -- ++(-13, +4) node[above right] {\footnotesize $\boldsymbol{b}_y$};
                    \draw[->, thick] (O) -- ++(0, -20) node[below] {\footnotesize $\boldsymbol{b}_x$};
                    \draw[->, thick] (O) -- ++(13, 5) node[above right] {\footnotesize $\boldsymbol{b}_z$};
                
                    % Motor locations
                    \coordinate (ML) at (15, 40);
                    \coordinate (MR) at (65, 40);
                
                    % Motor rotation arrows
                    % Left motor (CCW)
                    \begin{scope}[rotate=-100]
                        \draw[->, black]
                            (ML) ++(2,12)
                            arc[start angle=100,end angle=240,x radius=14,y radius=13];
                    \end{scope}

                    \node[left, yshift=-5pt, xshift=3pt] at (ML) {\footnotesize $\omega_2$};
                
                    % Right motor (CW)
                    \begin{scope}[rotate=-100]
                        \draw[->, black]
                            (MR) ++(14,-19)
                            arc[start angle=240,end angle=100,x radius=20,y radius=20];
                    \end{scope}
                    \node[right, yshift=-26pt, xshift=-13pt] at (MR) {\footnotesize $\omega_1$};

                    % Elevon locations (approximate)
                    \coordinate (EL) at (22, 18);  % left elevon
                    \coordinate (ER) at (58, 18);  % right elevon
                    
                    % Left elevon deflection (downward rotation)
                    \draw[->, black]
                        (EL) ++(-1,40)
                        arc[start angle=100,end angle=-30,x radius=3.5,y radius=5];
                    \node[left, yshift=45pt, xshift=4pt] at (EL) {\footnotesize $\delta_{2}$};
                    
                    % Right elevon deflection (downward rotation)
                    \draw[->, black]
                        (ER) ++(59,13)
                        arc[start angle=100,end angle=-30,x radius=7,y radius=10];
                    \node[right, yshift=10pt, xshift=65pt] at (ER) {\footnotesize $\delta_{1}$};

                \end{tikzpicture}

        \end{overpic}
        \caption{Cyclone in forward flight.}
        \label{photo_cyclone_fw}
    \end{subfigure}

    \caption{The Cyclone tailsitter \gls{uav}. Body reference frame and positive actuator input conventions.}
    \label{photo_cyclone}
\end{figure}

We consider the inertial reference frame $\mathcal{I}$, chosen as the 
Earth-fixed \gls{ned} frame with basis vectors 
$\{\boldsymbol{i}_x, \boldsymbol{i}_y, \boldsymbol{i}_z\}$, and the 
body-fixed reference frame $\mathcal{B}$ with basis vectors 
$\{\boldsymbol{b}_x, \boldsymbol{b}_y, \boldsymbol{b}_z\}$, depicted in \cref{photo_cyclone_fw}. The axis $\boldsymbol{b}_z$ coincides with the 
longitudinal axis of the vehicle and $\boldsymbol{b}_y$ with the lateral 
axis, while $\boldsymbol{b}_x$ is defined by the right-hand rule as 
$\boldsymbol{b}_x = \boldsymbol{b}_y \times \boldsymbol{b}_z$. The attitude of the vehicle is represented by the rotation matrix 
\begin{equation}
R_b^i = 
    \begin{bmatrix}
        \boldsymbol{b}_x & \boldsymbol{b}_y & \boldsymbol{b}_z
    \end{bmatrix}
    \in SO(3),
\end{equation}
which maps any vector expressed in the body-fixed frame to
the inertial according to
$
    \boldsymbol{x}^i = R_b^i \boldsymbol{x}^b
\footnote{Vector superscripts indicating the reference frame may be omitted if the meaning is clear from context.},$ where $\boldsymbol{x}$ is a generic vector. 
Equivalently, the attitude is represented by the normalized quaternion 
\begin{equation}
    \boldsymbol{q} = \begin{bmatrix}
        q_s & \boldsymbol{q}_v
    \end{bmatrix}^\top,
\end{equation}
where $q_s$ is the scalar part and $\boldsymbol{q}_v$ the vector part of the quaternion. The corresponding rotation of a vector can then be written as
$
    \boldsymbol{x}^b
    =
    \boldsymbol{q} \otimes 
    \begin{bmatrix}
        0 & \boldsymbol{x}^i
    \end{bmatrix}^\top
    \otimes \boldsymbol{q}^{*},
$
where $\boldsymbol{q}^*$ denotes the quaternion conjugate, and $\otimes$ the Hamilton product \cite{stevens-lewis}.

\subsection{Equations of motion}
The translational dynamics of the vehicle are given by
\begin{subequations}\label{eq:transl_dyn}
\begin{align}
    \dot{\boldsymbol{p}} &= \boldsymbol{v} \label{eq:xdot} \\
    \dot{\boldsymbol{v}}
        &= {R_b^i}\boldsymbol{f}^b(\boldsymbol{v}_a,\boldsymbol{u}) + \boldsymbol{g},
        \label{eq:vdot}
\end{align}
\end{subequations}
where $\boldsymbol{p}$ and $\boldsymbol{v}$ are, respectively, the position and the velocity of the vehicle in the inertial frame, and
\begin{equation}
    \boldsymbol{u} =
    \begin{bmatrix}
        \delta_1 & \delta_2 & \omega_1 & \omega_2
    \end{bmatrix}^\top,
\end{equation}
is the control input with $\delta_1$ and $\delta_2$ the deflection of the left and right elevon and $\omega_1$ and $\omega_2$ the rotational speed of the left and right propeller respectively. \Cref{photo_cyclone_fw} illustrates the conventions used for positive control inputs. The net specific force $\boldsymbol{f}$\footnote{Throughout the paper, notation is adapted to the context to improve readability; for instance, $\boldsymbol{f}(\boldsymbol{v}_a,\boldsymbol{u})$ is abbreviated as $\boldsymbol{f}$ when its dependencies are clear.} is a function of the velocity relative to the airmass $\boldsymbol{v_a}$ and the control input $\boldsymbol{u}$. The rotational dynamics are given by
\begin{subequations}\label{eq:rot_dyn}
\begin{align}
    \dot{\boldsymbol{q}} &= \frac{1}{2}\boldsymbol{q}\otimes\boldsymbol{\Omega} \\
    \dot{\boldsymbol{\Omega}}
        &= J^{-1}\boldsymbol{M}(\boldsymbol{v}_a,\boldsymbol{u}) - J^{-1}\boldsymbol{\Omega} \times J \boldsymbol{\Omega},
        \label{eq:Omegadot}
\end{align}
\end{subequations}
where $\boldsymbol{\Omega}$ is the angular rate of the vehicle, $\boldsymbol{M}$ is net moment as a function $\boldsymbol{v_a}$ and $\boldsymbol{u}$, and $J$ is the moment of inertia matrix. Finally, the relation between the vehicle velocity $\boldsymbol{v}$ and its velocity relative to the airmass $\boldsymbol{v}_a$ is
\begin{equation}
    \label{eq:va}
    \boldsymbol{v}_a = \boldsymbol{v} - \boldsymbol{v}_w,
\end{equation}
where $\boldsymbol{v}_w$ is the wind velocity.

\subsection{\texorpdfstring{$\phi$}{phi}-theory parameterization}
\label{subsec:model}
We employ the $\phi$-theory tailsitter model developed in \cite{lustosa2019} to express the net force and moment acting on the vehicle. This parameterization offers two key advantages over the standard Buckingham $\pi$-theory formulation \cite{anderson2016}. First, it provides a global model applicable across the entire flight envelope of the tailsitter, including transition phases and post-stall behavior, without requiring costly identification of lift and drag coefficients. Second, it avoids the singularities inherent to the Buckingham–$\pi$ formulation, present in near-hover conditions when the angle of attack $\alpha$ and sideslip $\beta$ are undefined. 

The target vehicle considered in this analysis is a modified version of the Cyclone tailsitter (\cref{photo_cyclone}), originally presented in \cite{bronz2017}. It is a typical tailsitter configuration featuring two fixed motors and two elevons. The modifications involve the use of a symmetric airfoil and alignment of the motor thrust axes with the longitudinal body axis. We adopt the body reference frame shown in the same figure, using hover-based conventions such that \(\boldsymbol{b}_x\) points forward during hover.

We utilize the \(\phi\)-theory formulation together with hover and forward-flight experimental data obtained from manual flights of the Cyclone to identify the model parameters. The net specific force in the body frame is modeled as
\begin{equation}
\label{eq:fb}
\boldsymbol{f}^b = \begin{bmatrix}
    f_{x}^b & f_{y}^b & f_{z}^b
\end{bmatrix}^\top,
\end{equation}
where
\begin{subequations}\label{eq:fbxyz}
\begin{align}
    f^b_{x} &=
        \begin{bmatrix}
            c_{x_1} & c_{x_2} & c_{x_3}
        \end{bmatrix}
        \begin{bmatrix}
            \|\boldsymbol{v}_a\| \boldsymbol{i}_x^\top \boldsymbol{v}_a^b \\
            \left(\delta_1 + \delta_2\right)\|\boldsymbol{v}_a\| \boldsymbol{i}_z^\top \boldsymbol{v}_a^b \\
            \delta_1\omega_1^2 + \delta_2\omega_2^2
        \end{bmatrix}
        \\
    f^b_{y} &= 
        c_{y_1}\,\|\boldsymbol{v}_a\| \boldsymbol{i}_y^\top \boldsymbol{v}_a^b
        \\
    f^b_{z} &=
        \begin{bmatrix}
            c_{z_1} & c_{z_2}
        \end{bmatrix}
        \begin{bmatrix}
            \|\boldsymbol{v}_a\| \boldsymbol{i}_z^\top \boldsymbol{v}_a^b \\
            \omega_1^2 + \omega_2^2
        \end{bmatrix}.
\end{align}
\end{subequations}
For the rotational dynamics we assume a diagonal moment of inertia matrix $J \in \mathbb{R}^{3 \times 3}
$, then if we set $\boldsymbol{m} = J^{-1}\boldsymbol{M}$ and $\boldsymbol{m}_\textrm{cor} = J^{-1}\boldsymbol{\Omega} \times J \boldsymbol{\Omega}$, we can write the equation \cref{eq:Omegadot} in the following compact form:
\begin{equation}
    \dot{\boldsymbol{\Omega}} = \boldsymbol{m} - \boldsymbol{m}_\textrm{cor}.
    \label{eq:mb}
\end{equation}
Applying the same reasoning as for the specific forces, we obtain the following formulation:
\begin{equation}
\boldsymbol{m} =
\begin{bmatrix}
    m_x & m_y & m_z
\end{bmatrix}^\top,
\end{equation}
where
\begin{subequations}\label{eq:mbxyz}
\begin{align}
    m_x &=
        \begin{bmatrix}
            \mu_{x_1} & \mu_{x_2}
        \end{bmatrix}
        \begin{bmatrix}
            \omega_1^2-\omega_2^2 \\
            \norm{\boldsymbol{v}_a}\boldsymbol{i}_y^\top \boldsymbol{v}_a^b \\
        \end{bmatrix}
        \\
    m_y &= 
        \begin{bmatrix}
            \mu_{y_1} & \mu_{y_2} & \mu_{y_3} & \mu_{y_4}
        \end{bmatrix}
        \begin{bmatrix}
            \delta_1\omega_1^2 + \delta_2\omega_2^2 \\
            \left(\delta_1+\delta_2\right)\norm{\boldsymbol{v}_a}\boldsymbol{i}_z^\top \boldsymbol{v}_a^b \\
            \norm{\boldsymbol{v}_a}\boldsymbol{i}_x^\top \boldsymbol{v}_a^b \\
            \ddot{\delta}_1 + \ddot{\delta}_2
        \end{bmatrix}
        \\
    m_z &=
        \begin{bmatrix}
            \mu_{z_1} & \mu_{z_2} & \mu_{z_3}
        \end{bmatrix}
        \begin{bmatrix}
            \delta_1\omega_1^2 - \delta_2\omega_2^2 \\
            \left(\delta_1-\delta_2\right)\norm{\boldsymbol{v}_a}\boldsymbol{i}_z^\top \boldsymbol{v}_a^b \\
            \omega_1^2 - \omega_2^2
        \end{bmatrix}.
\end{align}
\end{subequations}
For the identification of the parameters in formulation \cref{eq:mb}, explicit knowledge of the moment of inertia matrix is not required, as this is embedded within the $\mu$ coefficients. This enables model identification without relying on platform-specific parameters.

We note that the formulation in \cref{eq:fbxyz,eq:mbxyz} corresponds to a simplified representation of the full $\phi$-theory model. This simplification arises because, during the identification process, certain state-dependent terms were not sufficiently excited, despite attempts to induce them through manual maneuvers. To obtain a model with improved generalizability, small or inconsistent terms were therefore neglected, such as the moments induced by rotational rate terms. Nevertheless, the unmodelled state-dependent effects are comparatively slow changing relative to the actuator-related terms and are thus appropriately handled by the incremental controller presented later, as demonstrated in~\cite{tal2022}. \Cref{tab:cyclone_coeff} lists the identified coefficients and describes the physical meaning of the associated terms. We refer to the moment about the $\boldsymbol{b}_z$ axis as the yawing moment and to the moment about the $\boldsymbol{b}_x$ axis as the rolling moment. To avoid confusion, we note that the use of the hover-based body frame alters the naming of yawing and rolling moments compared to the standard aircraft body frame convention.

\begin{table}[h]
    \centering
    \caption{Cyclone model coefficients.}
    \begin{tabular}{c p{0.32\columnwidth} S[table-format=1.2e1] l}
        \toprule
        {Parameter} & {Physical meaning} & {Value} & {Units}\\
        \midrule
        $c_{x_1}$ & wing lift& -1.44e1 & $\mathrm{m^{-1}}$ \\ 
        $c_{x_2}$ & elevon-freestream lift& 6.83e-2 & $\mathrm{m^{-1}}$\\ 
        $c_{x_3}$ & elevon-propwash lift& -8.80e-6 & $\mathrm{m}$\\ 
        \midrule
        $c_{y_1}$ & lateral velocity drag & -8.41e-2 & $\mathrm{m^{-1}}$\\ 
        \midrule
        $c_{z_1}$ & drag & -3.00e-2 & $\mathrm{m^{-1}}$\\
        $c_{z_2}$ & thrust & -7.35e-6 & $\mathrm{m}$\\ 
        \midrule
        $\mu_{x_1}$ & differential thrust rolling moment & 3.90e-5 & --\\ 
        $\mu_{x_2}$ & lateral velocity rolling moment& 3.71e-3 & $\mathrm{m^{-2}}$\\ 
        \midrule
        $\mu_{y_1}$ & elevon-propwash pitching moment& -4.24e-5 & -- \\
        $\mu_{y_2}$ & elevon-freestream pitching moment & 2.53e-1 & $\mathrm{m^{-2}}$\\
        $\mu_{y_3}$ & vertical velocity pitching moment & -8.88e-2 & $\mathrm{m^{-2}}$\\
        $\mu_{y_4}$ & accelerating elevon mass pitching moment & -7.68e-3 & $\mathrm{s^{-2}}$ \\
        \midrule
        $\mu_{z_1}$ & elevon-propwash yawing moment & -5.83e-5 & -- \\
        $\mu_{z_2}$ & elevon-freestream yawing moment & 3.44e-1 & $\mathrm{m^{-2}}$\\ 
        $\mu_{z_3}$ & propeller reaction yawing moment & -3.97e-6 & -- \\ 
        \bottomrule
    \end{tabular}
    \label{tab:cyclone_coeff}
\end{table}

\section{DIFFERENTIAL FLATNESS UNDER WIND}
\label{sec:flatness}
We adopt the differential flatness transform introduced in \cite{tal2022} as the basis for the attitude and acceleration controller. In this section, we follow the methodology of the original work and present the resulting formulation, which incorporates
\begin{itemize}
    \item modifications to explicitly account for the vehicle velocity relative to the airmass $\boldsymbol{v}_a$ through \cref{eq:va}. 
    \item adaptations specific to our platform, namely a zero thrust angle, a symmetric airfoil, and a different body frame; these effects are captured by the tailsitter model developed in \cref{subsec:model}.
\end{itemize}
Derivations are omitted for brevity; readers are referred to \cite{tal2022} for details.

First, a simplified version of the tailsitter model is considered by neglecting the less dominant terms of the model such that the transform is feasible and simpler. These involve the state dependent terms $c_{y_1}$, $\mu_{x_2}$ and $\mu_{y_3}$, and the higher order actuator term $\mu_{y_4}$, which are set to zero.

Given the desired flat output\footnote{The flat output of the system is the desired, not the actual, signal. We do not explicitly use the subscript ``r" that commonly denotes a reference signal, since $\boldsymbol{p}$ is not necessarily obtained by a reference trajectory, as shown later. Thus we adopt a generic notation.} of the system 
\begin{equation}
\label{eq:flat_output}
\boldsymbol{\sigma} = \begin{bmatrix} \boldsymbol{p} & \psi \end{bmatrix}^\top
\end{equation}
and assuming $\boldsymbol{p}$ is at least four times differentiable and $\psi$ at least two \cite{tal2022}, we set $\dot{\boldsymbol{p}} = \boldsymbol{v}$, $\ddot{\boldsymbol{p}} = \boldsymbol{a}$, $\dddot{\boldsymbol{p}} = \boldsymbol{j}$. Subsequently, assuming known wind velocity $\boldsymbol{v}_w$, we obtain the velocity relative to the airmass using \cref{eq:va} and the specific force required by the vehicle with
\begin{equation}
\label{eq:f}
\boldsymbol{{f}} = \boldsymbol{a} - \boldsymbol{g}.
\end{equation}

\subsection{Attitude and specific thrust}
The attitude of the tailsitter is recovered from the flat output in Euler angles representation. A Z-X-Y rotation sequence is chosen, i.e., yaw \(\psi\), roll \(\phi\), and pitch \(\theta\). This choice is motivated by the fact that the singularity occurs at a roll angle of \( \phi = \pm 90^\circ \), rather than at a pitch angle of \( \theta = \pm 90^\circ \). This is appropriate for tailsitters, which operate over a large pitch angle range. It is noted that a rotation matrix will also be used next to express rotations between intermediate frames, e.g., $R_\phi^\theta$ represents the last rotation of the Z-X-Y sequence.

The roll angle $\phi$ of the tailsitter can be obtained by 
\begin{equation}
    \phi = -\textrm{atan2}\left(\beta_x,\beta_z\right) + k\pi, \\
    \label{eq:flatness:phi}
\end{equation}
where
\begin{subequations}
\begin{align}
    \beta_x &= \boldsymbol{i}_y^\top\boldsymbol{f}^\psi \\
    \beta_z &= \boldsymbol{i}^\top_z\boldsymbol{f}^i,
    \label{eq:flatness:bxbz}
\end{align}
\end{subequations}
and $\boldsymbol{f}^\psi = R_i^\psi\boldsymbol{f}^i$; $k \in \{0,1\}$ such that the obtained $R_\phi^i \boldsymbol{i}_y$ and the current $\boldsymbol{b}_y$ axes satisfy $\boldsymbol{b}_y \cdot (R_\phi^i \boldsymbol{i}_y) > 0$, where ($\cdot$) is the inner product of two vectors. 

The pitch angle $\theta$ and the specific thrust $\tau~=~c_{z_2}(\omega_1^2+\omega_2^2)$ are given by
\begin{equation}
    \theta = \textrm{atan2}(\sigma_x,\sigma_z) + k\pi \\
    \label{eq:flatness:theta}
\end{equation}
and
\begin{equation}
\begin{aligned}
\label{eq:flatness:T}
        \tau &= \sin \theta \boldsymbol{i}_x^\top \boldsymbol{f}^\phi + \cos \theta \boldsymbol{i}_z^\top \boldsymbol{f}^\phi \\ &- c_{z_1} \norm{\boldsymbol{v}_a}\big(\sin \theta \boldsymbol{i}_x^\top \boldsymbol{v}_a^\phi + \cos \theta \boldsymbol{i}_z^\top \boldsymbol{v}_a^\phi \big),
\end{aligned}
\end{equation}
respectively, where
\begin{subequations}
\begin{align}
\sigma_x &=
    \boldsymbol{i}_x^\top\boldsymbol{f}^\phi
    - c_{x_1}\norm{\boldsymbol{v}_a}\boldsymbol{i}_x^\top \boldsymbol{v}_a^\phi
    - c_{x_2} \Delta \norm{\boldsymbol{v}_a}\boldsymbol{i}_z^\top \boldsymbol{v}_a^\phi \notag \\
&\quad
    - \frac{c_{x_3}}{2c_{z_2}} \Delta
    \left(
        \boldsymbol{i}_z^\top \boldsymbol{f}^\phi
        - c_{z_1}\norm{\boldsymbol{v}_a}\boldsymbol{i}_z^\top \boldsymbol{v}_a^\phi
    \right) \\
\sigma_z &=
    \boldsymbol{i}_z^\top \boldsymbol{f}^\phi
    - c_{x_1}\norm{\boldsymbol{v}_a}\boldsymbol{i}_z^\top \boldsymbol{v}_a^\phi
    + c_{x_2} \Delta \norm{\boldsymbol{v}_a}\boldsymbol{i}_x^\top \boldsymbol{v}_a^\phi \notag\\
&\quad
    + \frac{c_{x_3}}{2c_{z_2}} \Delta
    \left(
        \boldsymbol{i}_x^\top \boldsymbol{f}^\phi
        - c_{z_1}\norm{\boldsymbol{v}_a}\boldsymbol{i}_x^\top \boldsymbol{v}_a^\phi
    \right),
\end{align}
\end{subequations}
and $\Delta = \delta_1 + \delta_2$ is assumed a known value, $\boldsymbol{v}_a^\phi = R_i^\phi \boldsymbol{v}_a^i$, $\boldsymbol{f}^\phi = R_i^\phi \boldsymbol{f}^i$, and $k \in \{0,1\}$ such that $\tau \geq 0$.

\subsection{Angular rate}
The angular rate of the vehicle can be recovered by 
\begin{equation}
    \boldsymbol{\Omega} = \begin{bmatrix} 0 \\ \dot{\theta} \\ 0 \end{bmatrix}
    + R_\phi^\theta \begin{bmatrix} \dot{\phi} \\ 0 \\ 0 \end{bmatrix}
    + R_\psi^\theta \begin{bmatrix} 0 \\ 0 \\ \dot{\psi} \end{bmatrix}.
    \label{eq:Wr}
\end{equation}
By differentiation of \cref{eq:flatness:phi}, the roll angle rate is obtained as
\begin{equation}
    \dot{\phi} =
    -\frac{\dot{\beta}_x\beta_z - \beta_x\dot{\beta}_z}{\beta_x^2+\beta_z^2},
\end{equation} where
\begin{subequations}
\begin{align}
\dot{\beta}_x &=
    - \cos\psi\,\dot{\psi}\,\boldsymbol{i}_x^\top \boldsymbol{f}^i
    - \sin\psi\, \boldsymbol{i}_x^\top \dot{\boldsymbol{f}}^i \notag\\
&\;\;\;\;
    - \sin\psi\,\dot{\psi}\,\boldsymbol{i}_y^\top \boldsymbol{f}^i
    + \cos\psi\, \boldsymbol{i}_y^\top \dot{\boldsymbol{f}}^i \\
\dot{\beta}_z &=
    \boldsymbol{i}_z^\top \dot{\boldsymbol{f}}^i
\end{align}
\end{subequations}
and by differentiation of \cref{eq:f}, $\dot{\boldsymbol{f}}^i = \boldsymbol{j}$. Taking the derivative of \cref{eq:flatness:theta} the pitch angle rate becomes
\begin{equation}
    \dot{\theta} =
    \frac{\dot{\sigma_x}\sigma_z - \sigma_x\dot{\sigma_z}}{\sigma_x^2+\sigma_z^2},
\end{equation} 
where
\begin{subequations}
\begin{align}
    \dot{\sigma}_x &= \boldsymbol{i}_x^\top \dot{\boldsymbol{f}}^\phi 
    - c_{x_1}\left(\norm{\dot{\boldsymbol{v}}_a} \boldsymbol{i}_x^\top \boldsymbol{v}_a^\phi +\norm{\boldsymbol{v}_a} \boldsymbol{i}_x^\top \dot{\boldsymbol{v}}_a^\phi \right) \notag\\
    &\quad- c_{x_2}\Delta\left(\norm{\dot{\boldsymbol{v}}_a} \boldsymbol{i}_z^\top \boldsymbol{v}_a^\phi +\norm{\boldsymbol{v}_a}\boldsymbol{i}_z^\top \dot{\boldsymbol{v}}_a^\phi\right)\notag\\ 
    &\quad- c_{x_3}\frac{1}{2c_{z_2}}\Delta\left(\boldsymbol{i}_z^\top \dot{\boldsymbol{f}}^\phi - c_{z_1}\left(\norm{\dot{\boldsymbol{v}}_a} \boldsymbol{i}_z^\top \boldsymbol{v}_a^\phi +\norm{\boldsymbol{v}_a}\boldsymbol{i}_z^\top \dot{\boldsymbol{v}}_a^\phi\right)\right) \\
    \dot{\sigma}_z &= \boldsymbol{i}_z^\top \dot{\boldsymbol{f}}^\phi 
    - c_{x_1}\left(\norm{\dot{\boldsymbol{v}}_a} \boldsymbol{i}_z^\top \boldsymbol{v}_a^\phi +\norm{\boldsymbol{v}_a}\boldsymbol{i}_z^\top \dot{\boldsymbol{v}}_a^\phi\right) \notag\\
    &\quad+ c_{x_2}\Delta\left(\norm{\dot{\boldsymbol{v}}_a}\boldsymbol{i}_x^\top \boldsymbol{v}_a^\phi +\norm{\boldsymbol{v}_a}\boldsymbol{i}_x^\top \dot{\boldsymbol{v}}_a^\phi\right)\notag\\ 
    &\quad+ c_{x_3}\frac{1}{2c_{z_2}}\Delta\left(\boldsymbol{i}_x^\top \dot{\boldsymbol{f}}^\phi - c_{z_1}\left(\norm{\dot{\boldsymbol{v}}_a}\boldsymbol{i}_x^\top \boldsymbol{v}_a^\phi +\norm{\boldsymbol{v}_a}\boldsymbol{i}_x^\top \dot{\boldsymbol{v}}_a^\phi\right)\right),
\end{align}
\end{subequations}
and
\begin{subequations}
    \begin{align}
    \dot{\boldsymbol{f}}^\phi &= \dot{R}_i^\phi \boldsymbol{f}^i + R_i^\phi \dot{\boldsymbol{f}}^i \\
        \dot{\boldsymbol{v}}_a^\phi &= \dot{R}_i^\phi \boldsymbol{v}_a^i + R_i^\phi \dot{\boldsymbol{v}}_a \\
        \norm{\dot{\boldsymbol{v}}_a} &= \frac{\boldsymbol{v}_a^\top \dot{\boldsymbol{v}}_a}{\norm{\boldsymbol{v}_a}} \\
        \dot{\boldsymbol{v}}_a &= \boldsymbol{a}-\dot{{\boldsymbol{v}}}_w.
    \end{align}
\end{subequations}
The yaw angle rate $\dot{\psi}$ is obtained trivially by \cref{eq:flat_output}.

We provide the full mathematical expressions; however, in reality, one may assume constant or slow changing wind ${\boldsymbol{v}}_w$, allowing the equations to be simplified by setting $\dot{{\boldsymbol{v}}}_w = 0$.

\subsection{Control input}
\label{subsec:flatness_actuators}
Finally, to recover the control input we first compute the rotational speed of the propellers with
\begin{equation}\label{eq:flatness:omegas}
    \begin{alignedat}{2}
    \omega_1 &= \sqrt{\frac{\varepsilon_1 + \varepsilon_2}{2}}
    &\qquad
    \omega_2 &= \sqrt{\frac{\varepsilon_1 - \varepsilon_2}{2}},
    \end{alignedat}
\end{equation}
where $\varepsilon_1 = \omega_1^2 + \omega_2^2 = \tau/c_{z_2}$ and $\varepsilon_2 = \omega_1^2 - \omega_2^2 = \boldsymbol{i}_x^\top \boldsymbol{m}/\mu_{x_1}$. Then we calculate the elevon deflections by
\begin{equation} \label{eq:flatness:deltas}
\begin{alignedat}{2}
    \delta_1 &= \frac{1}{\zeta_1} \left( \eta_z-\zeta_2\delta_2 \right)
    &\qquad
    \delta_2 &= \frac{\eta_y-\frac{\zeta_3}{\zeta_1}\eta_z}{\zeta_4-\frac{\zeta_3}{\zeta_1}\zeta_2}, 
\end{alignedat}
\end{equation}
where
\begin{subequations}
\begin{align}
    \eta_y &= \boldsymbol{i}_y^\top \boldsymbol{m}\\
    \eta_z &= \boldsymbol{i}_z^\top \boldsymbol{m} - \mu_{z_3}(\omega_1^2 - \omega_2^2)\\
    \zeta_1 &= \mu_{z_1}\omega_1^2 + \mu_{z_2}\norm{\boldsymbol{v}_a}\boldsymbol{i}_z^\top \boldsymbol{v}_a^b \\
    \zeta_2 &= -\mu_{z_1}\omega_2^2 - \mu_{z_2}\norm{\boldsymbol{v}_a}\boldsymbol{i}_z^\top \boldsymbol{v}_a^b \\
    \zeta_3 &= \mu_{y_1}\omega_1^2 + \mu_{y_2}\norm{\boldsymbol{v}_a}\boldsymbol{i}_z^\top \boldsymbol{v}_a^b \\
    \zeta_4 &= \mu_{y_1}\omega_2^2 + \mu_{y_2}\norm{\boldsymbol{v}_a}\boldsymbol{i}_z^\top \boldsymbol{v}_a^b.
\end{align}
\end{subequations}

\section{CONTROLLER DESIGN}
\subsection{Acceleration and attitude controller}
\label{subsec:ll_controller}
We use the \gls{indi} controller structure for the acceleration and the angular acceleration controllers as originaly developed in \cite{tal2022}. For completeness we briefly present them in the following. The incremental control law for the acceleration controller is 
\begin{equation}
    \label{eq:indi_accel}
    \boldsymbol{f}_c^i = (\boldsymbol{a}_c - \boldsymbol{a}_{f}) + \boldsymbol{f}_f^i,
\end{equation}
where $\boldsymbol{a}_c$ denotes the input signal provided by the higher-level guidance controller, $\boldsymbol{a}_f$ is the low-pass filtered acceleration feedback, and $\boldsymbol{f}_f^i$ is obtained from \cref{eq:fb} using the filtered control input signal. Subsequently, $\boldsymbol{f}_c^i$ is used with \cref{eq:flatness:phi,eq:flatness:theta} to compute the attitude control signal $\boldsymbol{q}_c$. The quaternion error is obtained with $\boldsymbol{q}_e~=~\boldsymbol{q}_c \otimes \boldsymbol{q}^*$ \cite{fresk2013}.
The following PD control law is used the for attitude control:
\begin{equation}
    \label{eq:att_controller}
    \dot{\boldsymbol{\Omega}}_c = -K_q\boldsymbol{q}_{e,v}
    -
    K_\Omega(\boldsymbol{\Omega} - \boldsymbol{\Omega}_\textrm{ff}),
\end{equation}
where $\boldsymbol{q}_{e,v}$ is the vector part of the quaterion error, $K_q$,~$K_\Omega~\in~\mathbb{R}^{3\times3}$ are diagonal gain matrices with positive elements, and $\boldsymbol{\Omega}_\textrm{ff}$ is the feedforward angular rate signal computed by \cref{eq:Wr}. Finally, the incremental angular acceleration control law is
\begin{equation}
    \boldsymbol{m}_c = (\dot{\boldsymbol{\Omega}}_c - \dot{\boldsymbol{\Omega}}_f) + \boldsymbol{m}_f,
\end{equation}
where $\dot{\boldsymbol{\Omega}}_f$ is the low-pass filtered angular acceleration feedback, and $\boldsymbol{m}_f$ is obtained from \cref{eq:mb} using the filtered control input signal. Then $\boldsymbol{m}_c$ is used with \cref{eq:flatness:omegas,eq:flatness:deltas} to calculate the control input sent to the actuators.

\subsection{GVF-based guidance controller}
\label{subsec:gvfcontroller}
In this subsection, we develop a GVF-based guidance controller to generate the signal $\boldsymbol{a}_c$ introduced in \cref{subsec:ll_controller}, building on and extending the parametric \gls{gvf} formulation of \cite{yao2021}. First, we define the desired path in the parameterized form, $\boldsymbol{p}_g = \boldsymbol{f}(w) \in \mathbb{R}^3$, where $w$ is the scalar path parameter. The point $\boldsymbol{p}_g$ is also denoted as the \textit{guiding point}. The position of the \gls{uav} is $\boldsymbol{p}\in \mathbb{R}^3$. We define the augmented position as $\boldsymbol{\xi} = \begin{bmatrix}\boldsymbol{p} & w\end{bmatrix}^\top$ and the path tracking error $\boldsymbol{\phi}(\boldsymbol{\xi}) = \boldsymbol{p} - \boldsymbol{p}_g$. We propose the following extended form:
\begin{equation}
    \boldsymbol{\chi}(\boldsymbol{\xi}) = \begin{bmatrix}
        \boldsymbol{\chi}_p\left(\boldsymbol{\xi}\right) \\ \chi_w\left(\boldsymbol{\xi}\right) 
    \end{bmatrix}
              = \begin{bmatrix}
            -\boldsymbol{f}' +\boldsymbol{f}'(K\boldsymbol{\phi})^\top\boldsymbol{f}'- K\boldsymbol{\phi} \\
            -1 + \left(K\boldsymbol{\phi}\right)^\top\boldsymbol{f}' 
    \end{bmatrix},
    \label{eq:ik-gvf}
\end{equation}
where $K\in \mathbb{R}^{3\times3}$ is a diagonal gain matrix with positive elements and $\boldsymbol{f}' = \frac{d\boldsymbol{f}}{dw}$. Following, we develop the guidance control law based on the proposed \gls{gvf} and subsequently justify the proposed modification to the \gls{gvf} structure by proving the exponential stability of the tracking error dynamics for a single-integrator system.

\subsubsection{Guidance controller}
To design the guidance controller based on the \gls{gvf} in \cref{eq:ik-gvf} we assume that the \gls{uav} is a second order system $\ddot{\boldsymbol{p}} =\boldsymbol{a}_c$, a standard approach widely adopted in the literature (\cite{tal2022}, \cite{ewoud2018}, \cite{ewoud2020}); $\boldsymbol{a}_c$ is the control input to design. We define the augmented velocity tracking error as the difference between the derivative of the augmented position $\dot{\boldsymbol{\xi}}$ and the normalized \gls{gvf} velocity vector $\boldsymbol{\chi}/\norm{\boldsymbol{\chi}_p}$ scaled by a reference speed $s_r \in \mathbb{R}$ as
\begin{equation} 
    \boldsymbol{e}_{\dot{\xi}} = \dot{\boldsymbol{\xi}} - s_r\frac{\boldsymbol{\chi}}{\norm{\boldsymbol{\chi}_p}} = \begin{bmatrix}
        \dot{\boldsymbol{p}} \\ \dot{w}
    \end{bmatrix} - \begin{bmatrix}
        \hat{s}_r\boldsymbol{\chi}_p \\ \hat{s}_r\chi_w
    \end{bmatrix},
\end{equation}
where $\hat{s}_r = s_r/\norm{\boldsymbol{\chi}_p}$. By choosing 
\begin{equation}
\dot{w} = \hat{s}_r\chi_w,
\label{eq:wdot}
\end{equation}
it suffices to design the control law such that the velocity tracking error $\boldsymbol{e}_v = \dot{\boldsymbol{p}} - \hat{s}_r\boldsymbol{\chi}_p$ converges to zero. We select the Lyapunov candidate function $V = \frac{1}{2}\boldsymbol{e_v}^\top\boldsymbol{e_v}$; its time derivative is
\begin{equation}
    \dot{V} = \dot{\boldsymbol{e}_v}^\top\boldsymbol{e}_v = \left(\boldsymbol{a}_c - \dot{\hat{s}}_r\boldsymbol{\chi}_p -\hat{s}_r\dot{\boldsymbol{\chi}}_p\right)^\top\left(\dot{\boldsymbol{p}} - \hat{s}_r\boldsymbol{\chi}_p\right),
\end{equation}
where 
\begin{equation}
\dot{\boldsymbol{\chi}}_p = \mathbf{J}_{\chi_p,p}\dot{\boldsymbol{p}} + \frac{\partial \boldsymbol{\chi}_p}{\partial w}\dot{w}.
\end{equation}
Notation $\mathbf{J}_{x,y}$ denotes the Jacobian of the generic vector $\boldsymbol{x}$ w.r.t. generic vector $\boldsymbol{y}$. We choose the control law 
\begin{equation}
\boldsymbol{a}_c = \dot{\hat{s}}_r\boldsymbol{\chi}_p + \hat{s}_r\dot{\boldsymbol{\chi}}_p - K_v(\dot{\boldsymbol{p}} - \hat{s}_r\boldsymbol{\chi}_p),
\end{equation}
where $K_v\in \mathbb{R}^{3\times3}$ is a diagonal gain matrix with positive elements. This results in $\dot{V} = -K_v\norm{\boldsymbol{e}_v}^2 \leq 0$, which means that, for any non-zero velocity error, $\boldsymbol{e}_v$ converges exponentially to zero. This is a PD control law as can be seen in the following compact form
\begin{equation}
    \boldsymbol{a}_c = \dot{\boldsymbol{v}}_c - K_v\left(\dot{\boldsymbol{p}}-\boldsymbol{v}_c\right),
    \label{eq:ac_gvf}
\end{equation}
where $\boldsymbol{v}_c = \hat{s}_r \boldsymbol{\chi}_p$, and $\dot{\boldsymbol{v}}_c = \dot{\hat{s}}_r\boldsymbol{\chi}_p + \hat{s}_r\dot{\boldsymbol{\chi}}_p$.

\subsubsection{Stability analysis of proposed GVF}
\label{subsubsec:gvftuning}
Here, we justify the use of the proposed extended parametric \gls{gvf} defined in \cref{eq:ik-gvf} by proving the stability of the system driven by this vector field. To this end, we consider the case of a single integrator system that tracks this \gls{gvf}, $\dot{\boldsymbol{p}} = \hat{s}_r\boldsymbol{\chi}_p$. This is a typical assumption used in related works \cite{yuri2018,bautista2025} and can be satisfied by tuning the inner velocity controller to be sufficiently faster than the GVF effective gain introduced below. Tuning is further discussed in \cref{subsec:simulation}. Expanding the relation for the single integrator and using \cref{eq:ik-gvf} we get the dynamics of the tracking error
\begin{equation}
\begin{aligned}
   \dot{\boldsymbol{p}} - \hat{s}_r\left( -\boldsymbol{f}' +\boldsymbol{f}'(K\boldsymbol{\phi})^\top\boldsymbol{f}'\right)-\hat{s}_r(-K\boldsymbol{\phi})&= 0 \Rightarrow \\
   \dot{\boldsymbol{p}} - \dot{\boldsymbol{p}_g} + \hat{s}_rK\boldsymbol{\phi}&= 0\Rightarrow \\
   \dot{\boldsymbol{\phi}} + K_\textrm{eff}\boldsymbol{\phi} &= 0.
\end{aligned}
\end{equation}
For the above derivation, the following relations are used, which are directly derived:
$\dot{\boldsymbol{p}}_g = \boldsymbol{f}'\dot{w}$ and $\dot{\boldsymbol{\phi}} = \dot{\boldsymbol{p}} - \dot{\boldsymbol{p}}_g$, together with \cref{eq:wdot}. In addition, the effective gain is defined as $K_{\textrm{eff}} = \hat{s}_r K$. Then the tracking error converges exponentially to zero for $K_\textrm{eff} > 0$. As such, we design the \gls{gvf} gain K to be
\begin{equation}
    K = {\hat{s}_r}^{-1}K_\textrm{eff}.
\end{equation}
This result enables the consistent tuning and comparison of the two guidance strategies, which is analyzed subsequently in \cref{sec:simulation}.

\subsection{Angular rate feedforward signal}
To compute the feedforward angular rate signal $\boldsymbol{\Omega}_{\mathrm{ff}}$ using \cref{eq:Wr}, the feedforward signals of velocity, acceleration, and jerk are required. 
In the case of a trajectory-tracking controller, these quantities are obtained as time derivatives of the reference trajectory. In contrast, within the \gls{gvf} guidance framework, the feedforward signals are position-dependent rather than time-dependent, and the guiding point is the equivalent of the reference point of a trajectory-tracking controller, as discussed in~\cite{yao2021}. Accordingly, the velocity, acceleration, and jerk of the guidance point are used in the computation of $\boldsymbol{\Omega}_{\mathrm{ff}}$.

We first observe that $\boldsymbol{\chi}_p$ from \cref{eq:ik-gvf} can be written in the following compact form:
\begin{equation}
   \boldsymbol{\chi}_p = \boldsymbol{\chi}_{p_g}- K\boldsymbol{\phi},
\end{equation}
where $\boldsymbol{\chi}_{p_g} = -\boldsymbol{f}'+\boldsymbol{f}'(K\boldsymbol{\phi})^\top\boldsymbol{f}'$. Then the guidance point's position derivatives are computed as follows:
\begin{subequations}
\begin{align}
\dot{\boldsymbol{p}_g} &= \hat{s}_r\boldsymbol{\chi}_{p_g}\\
        \ddot{\boldsymbol{p}}_g &= \dot{\hat{s}}_r\boldsymbol{\chi}_{p_g} +\hat{s}_r \dot{\boldsymbol{\chi}}_{p_g}\\
        \dddot{\boldsymbol{p}}_g &= \ddot{\hat{s}}_r\boldsymbol{\chi}_{p_g} + 2\dot{\hat{s}}_r\dot{\boldsymbol{\chi}}_{p_g} + \hat{s}_r\ddot{\boldsymbol{\chi}}_{p_g},
\label{eq:pg_derivatives}
\end{align}
\end{subequations}
where
\begin{subequations}
\begin{align}
\dot{\boldsymbol{\chi}}_{p_g} &= \mathbf{J}_{\chi_{p_g},p}\dot{\boldsymbol{p}} + \frac{\partial \boldsymbol{\chi}_{p_g}}{\partial w}\dot{w}
\\
    \ddot{\boldsymbol{\chi}}_{p_g} &= 
    \mathbf{J}_{\chi_{p_g},p}\ddot{\boldsymbol{p}} + 
     2\mathbf{H}_{\chi_{p_g},pw}\dot{\boldsymbol{p}}\dot{w}+
    \frac{\partial \boldsymbol{\chi}_{p_g}}{\partial w}\ddot{w}+
    \frac{\partial^2\boldsymbol{\chi}_{p_g}}{\partial w^2}\dot{w}^2,
\end{align}
\end{subequations}
and 
\begin{subequations}
    \begin{align}
        \dot{\hat{s}}_r &= \dot{s}_r\frac{1}{\norm{{\boldsymbol\chi}_p}} -s_r\frac{\dot{\boldsymbol{\chi}}_p^\top\boldsymbol{\chi}_p}{\norm{{\boldsymbol\chi}_p}^3}\\
        \ddot{\hat{s}}_r &= \ddot{s}_r\frac{1}{\norm{{\boldsymbol\chi}_p}} -2\dot{s}_r\frac{\dot{\boldsymbol{\chi}}_p^\top\boldsymbol{\chi}_p}{\norm{{\boldsymbol\chi}_p}^3}
        -s_r\frac{\ddot{\boldsymbol{\chi}}_p^\top\boldsymbol{\chi}_p + \dot{\boldsymbol{\chi}}_p^\top\dot{\boldsymbol{\chi}}_p}{\norm{{\boldsymbol\chi}_p}^3} \notag\\&\quad+3s_r\frac{(\dot{\boldsymbol{\chi}}_p^\top\boldsymbol{\chi}_p)^2}{\norm{{\boldsymbol\chi}_p}^5}.
    \end{align}
\end{subequations}
The second derivative of $w$ is found as
\begin{equation}
    \ddot{w} = \dot{\hat{s}}_r\chi_w + \hat{s}_r\dot{\chi}_w,
\end{equation}
where 
\begin{equation}
    \dot{\chi}_w = \mathbf{J}_{\chi_w,p}\dot{p} + \frac{\partial \chi_{w}}{\partial w}\ddot{w}.
\end{equation}
Notation $\mathbf{H}_{x,yz}$ denotes the Hessian matrix of the generic vector $\boldsymbol{x}$ w.r.t. the generic vectors $\boldsymbol{y}$ and $\boldsymbol{z}$. The second derivative of $\boldsymbol{\chi}_p$ is computed by
\begin{equation}
    \ddot{\boldsymbol{\chi}}_{p} =
    \mathbf{J}_{\chi_{p},p}\ddot{\boldsymbol{p}} + 
    2\mathbf{H}_{\chi_{p},pw}\dot{\boldsymbol{p}}\dot{w}+
    \frac{\partial \boldsymbol{\chi}_{p}}{\partial w}\ddot{w}+
    \frac{\partial^2\boldsymbol{\chi}_{p}}{\partial w^2}\dot{w}^2.
\end{equation}
Finally, we compute the feedforward angular rate signal $\boldsymbol{\Omega}_\textrm{ff}$ by \cref{eq:Wr,eq:pg_derivatives}. Detailed matrix expressions are given in the Appendix.

\section{SIMULATION STUDY}
\label{sec:simulation}

\subsection{Simulation model}
\begin{figure}[h]
    \centering
    \includegraphics[width=0.8\columnwidth]{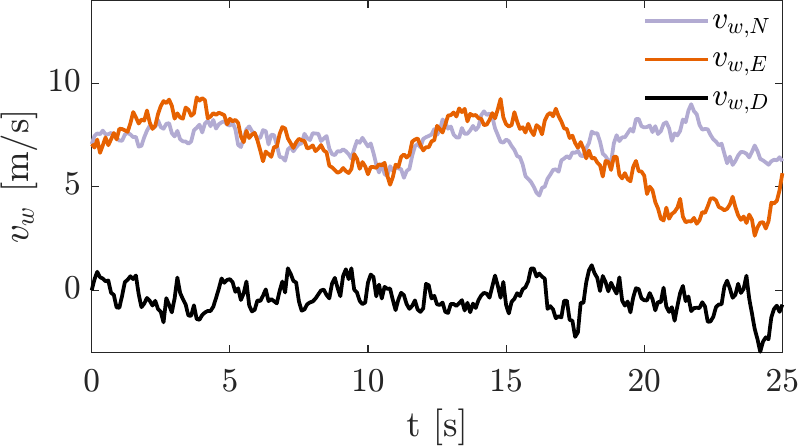}
    \caption{Wind velocity experienced by the \gls{uav} as it traverses a frozen turbulence field at speed $s_r$, shown for a southwest wind with $\|\boldsymbol{v}_{w,s} \| = 10~\mathrm{m/s}$ in the \gls{ned} frame.}
    \label{fig:wind}
\end{figure}

We employ the $\phi$-theory model presented in \cref{sec:model} to describe the tailsitter dynamics. The wind velocity $\boldsymbol{v}_w$ introduced in \cref{sec:flatness} is formulated as 
\begin{equation}
   \label{eq:wind}
   \boldsymbol{v}_w = \boldsymbol{v}_{w,s} + \boldsymbol{v}_{w,g},
\end{equation}
where $\boldsymbol{v}_{w,s}$ is the mean wind velocity, assumed constant or slow changing, and $\boldsymbol{v}_{w,g}$ is the wind gust velocity. 
%Further we assume an estimator that produces $\boldsymbol{\hat v}_w$ estimates of the wind. 
The mean wind velocity $\boldsymbol{v}_{w,s}$ is specified as an input to the simulation, while the corresponding wind gust component $\boldsymbol{v}_{w,g}$ is generated using the Dryden turbulence model in accordance with the MIL-HDBK-1797 standard for low altitude, following the implementation described in \cite{hakim2018}. Wind effects are modeled accordingly in \cref{eq:fb,eq:mb}. As an example, \cref{fig:wind} depicts the wind velocity used in the simulations for the case of southwest wind with $\| \boldsymbol{v}_{w,s} \| = 10~\mathrm{m/s}$. 

\subsection{Comparison and tuning of guidance control strategies}
\label{subsec:simulation}
The trajectory-tracking guidance controller used in \cite{tal2022} is a simple PD controller given by
\begin{equation}
\label{eq:ac_traj}
    \boldsymbol{a}_c = \boldsymbol{a}_r -K_v'(\boldsymbol{v} - \boldsymbol{v}_r) - K_p'(\boldsymbol{p} - \boldsymbol{p}_r),
\end{equation}
where $\boldsymbol{p}_r(t) \in \mathbb{R}^{3}$ denotes the reference trajectory, $\boldsymbol{v}_r = \dot{\boldsymbol{p}}_r$, and $\boldsymbol{a}_r = \ddot{\boldsymbol{p}}_r$. Using \cref{eq:Wr} and $\boldsymbol{j}_r = \dddot{\boldsymbol{p}}_r$, the authors compute the feedforward angular rate signal $\boldsymbol{\Omega}_\textrm{ff}$, purely based on reference signals. The acceleration and attitude control structure is the same as described in \cref{subsec:ll_controller}.

The objective is to tune the above trajectory controller such that the \gls{uav} follows the reference trajectory point $\boldsymbol{p}_r$ with the same dynamics as those obtained when the \gls{uav} follows the guiding point $\boldsymbol{p}_g$ under the proposed \gls{gvf}-based guidance controller. To this end, we first rewrite the \gls{gvf} guidance law found in \cref{subsec:gvfcontroller} as
\begin{subequations}
    \begin{align}
        \boldsymbol{a}_c & = \dot{\boldsymbol{v}}_c - K_v(\boldsymbol{\dot{p}} - \boldsymbol{v}_c) \\
        \boldsymbol{v}_c &= \dot{\boldsymbol{p}}_g - K_\textrm{eff}(\boldsymbol{p} - \boldsymbol{p}_g),
    \end{align}
\end{subequations}
which can be expressed equivalently as
\begin{equation}
    \boldsymbol{a}_c = \ddot{\boldsymbol{p}}_g -(K_v + K_\textrm{eff})(\dot{\boldsymbol{p}} - \dot{\boldsymbol{p}}_g) - K_\textrm{eff}K_v(\boldsymbol{p} - \boldsymbol{p}_g).
\end{equation}
By direct comparison with \cref{eq:ac_traj}, the trajectory guidance controller gains can therefore be selected as
\begin{subequations}
\begin{align}
K_p' &= K_\textrm{eff}K_v \\
    K_v' &= K_v + K_\textrm{eff}.
\end{align}
\end{subequations}
For the \gls{gvf} guidance the tracking dynamics gain is set to $K_{\mathrm{eff}} = 0.5$. 
To satisfy the first-order kinematics assumption, a sufficiently larger gain is selected for the velocity controller, $K_v = 5$. 
Under these settings, the equivalent tuning for the trajectory-guidance scheme is $K_p' = 2.5$ and $K_v' = 5.5$.

\subsection{Yaw angle}
Assuming coordinated and level flight, the resulting yaw angle and its analytical derivative, which are used as inputs to the differential flatness transform, are computed as follows
\begin{subequations}
\begin{align}    
    \psi &= \textrm{atan2}\left(\boldsymbol{i}_y^\top \boldsymbol{v}_a, \boldsymbol{i}_x^\top \boldsymbol{v}_a\right) \\
    \dot{\psi} &= \frac{\boldsymbol{i}_y^\top \dot{\boldsymbol{v}}_a \boldsymbol{i}_x^\top \boldsymbol{v}_a - \boldsymbol{i}_y^\top \boldsymbol{v}_a \boldsymbol{i}_x^\top \dot{\boldsymbol{v}}_a}  {\left(\boldsymbol{i}_x^\top \boldsymbol{v}_a\right)^2 + \left(\boldsymbol{i}_y^\top \boldsymbol{v}_a\right)^2}.
\end{align}
\end{subequations}

\subsection{Nominal performance}
First, we simulate the system without wind. The purpose is to assess the nominal performance of both strategies in an ideal scenario without disturbances. We consider two representative maneuvers parameterized by $w$ for the \gls{gvf}-based guidance: a level circular path centered at $(0,0)$,
\begin{equation}
\boldsymbol{f}^{\textrm{circ}}(w)=
\begin{bmatrix}
    r\cos w\\
    r\sin w\\
    z_0
\end{bmatrix},
\end{equation}
with $r = 20\,\unit{\meter}$ and arbitrary $z_0$, and a Lissajous curve,
\begin{equation}
\boldsymbol{f}^{\textrm{liss}}(w)=
\begin{bmatrix}
    c_x \cos(\omega_x w + d_x)\\
    c_y \cos(\omega_y w + d_y)\\
    c_z \cos(\omega_z w + d_z)
\end{bmatrix},
\end{equation}
with $c_x = 50\,\unit{\meter}$, $c_y = 15\,\unit{\meter}$, $c_z = 5\,\unit{\meter}$, $\omega_x = 1$, $\omega_y = 2$, $\omega_z = 2$, $d_x = 0$, $d_y = \pi/2$, and $d_z = 0$.

\begin{figure}[h]
    \centering
    \includegraphics[width=0.8\columnwidth]{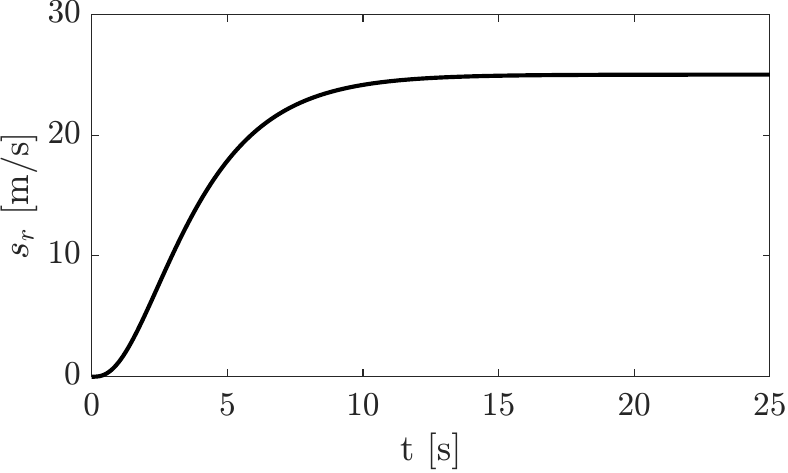}
    \caption{Reference speed $s_r$ with a final value of $V_s = 25~\mathrm{m/s}$.}
    \label{fig:sr_25}
\end{figure}

The reference speed $s_r$ is obtained from the reference final speed $V_s$, which is provided as input to a third-order transfer function $G(s)$, yielding a smooth transient response compatible with the differential flatness requirements on the position derivatives. Thus
\begin{equation}
s_r(s) = V_sH(s)G(s),
\end{equation}
where $H(s)$ is the heavyside function and $G(s)~=~\omega_n^2 / \bigl( (s^2 + 2\zeta \omega_n s + \omega_n^2)(\alpha s + 1) \bigr)$. The parameters are selected as $\zeta = 1.2$, $\omega_n = 0.8$, and $\alpha = 1.25$. \Cref{fig:sr_25} shows an example reference speed $s_r$ with a final value of $V_s =25~\mathrm{m/s}$.

The corresponding reference trajectory $\boldsymbol{p}_r$ for the trajectory-guidance controller, subject to the constraint $s_r$, can be derived numerically offline using arc-length parameterization.

\subsubsection{Without initial error}
\label{subsubsec:no_initial_error}
In this scenario the \gls{uav} is positioned initially on the curve. We expect comparable performance from the two guidance strategies. 
The principal difference of the \gls{gvf} guidance is that the parameter $w$, which parameterizes the desired curve, is not evolving in open loop like time, but rather in closed loop with the state of the system. Since no disturbances are considered other than the assumptions introduced by the differential flatness transform and the controller design, no significant performance difference is expected. Under these conditions, the \gls{gvf} guidance behaves similarly to a trajectory-tracking controller, as explicitly discussed in~\cite{yao2021}.

\begin{figure*}[h]
    \centering
    \begin{subfigure}[t]{\columnwidth}
        \centering
        \includegraphics[width=0.8\columnwidth]{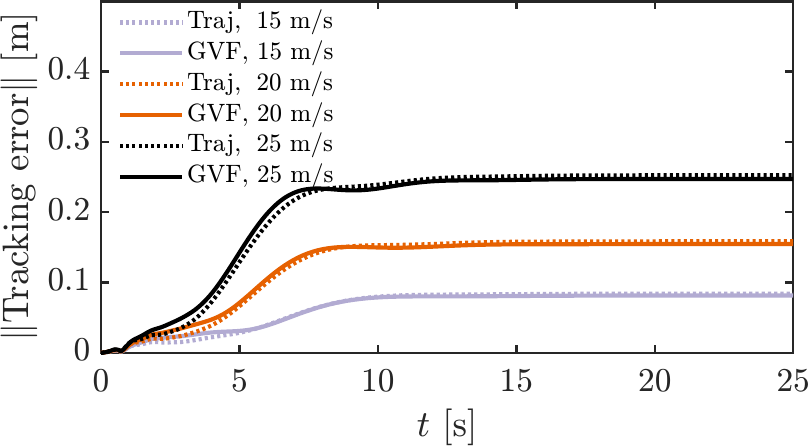}
        \caption{$\| \boldsymbol{v}_{w,s} \|= 0~\mathrm{m/s}$.}
        \label{fig:circle_nowind}
    \end{subfigure}
    \begin{subfigure}[t]{\columnwidth}
        \centering
        \includegraphics[width=0.8\columnwidth]{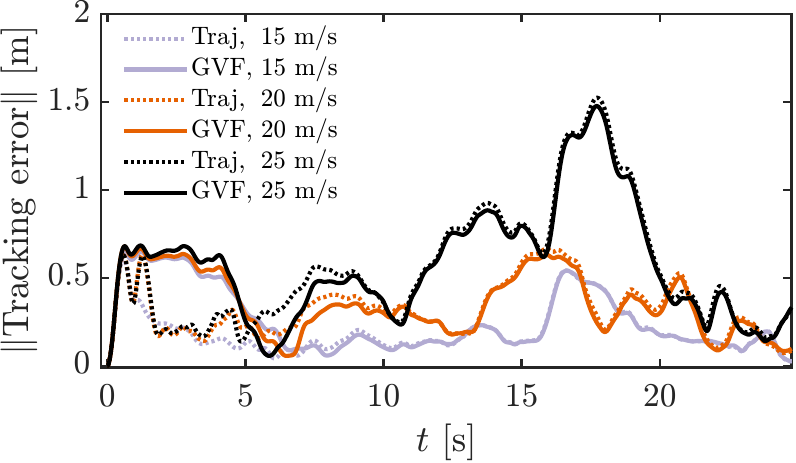}
        \caption{$\| \boldsymbol{v}_{w,s} \| = 10~\mathrm{m/s}$.}
        \label{fig:circle_wind}
    \end{subfigure}

    \caption{Tracking error norm on the circular curve. Zero-wind and $\| \boldsymbol{v}_{w,s}\|~=10~\textrm{m/s}$ case.}
    \label{fig:circle}
\end{figure*}

\begin{figure*}[h]
    \centering
    \begin{subfigure}[t]{\columnwidth}
        \centering
        \includegraphics[width=0.8\columnwidth]{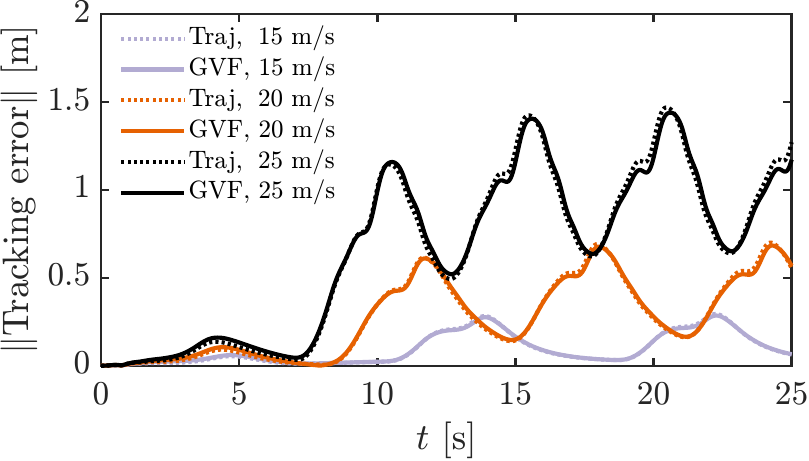}
        \caption{$\| \boldsymbol{v}_{w,s} \| = 0~\mathrm{m/s}$.}
        \label{fig:lissajous_nowind}
    \end{subfigure}
    \begin{subfigure}[t]{\columnwidth}
        \centering
        \includegraphics[width=0.8\columnwidth]{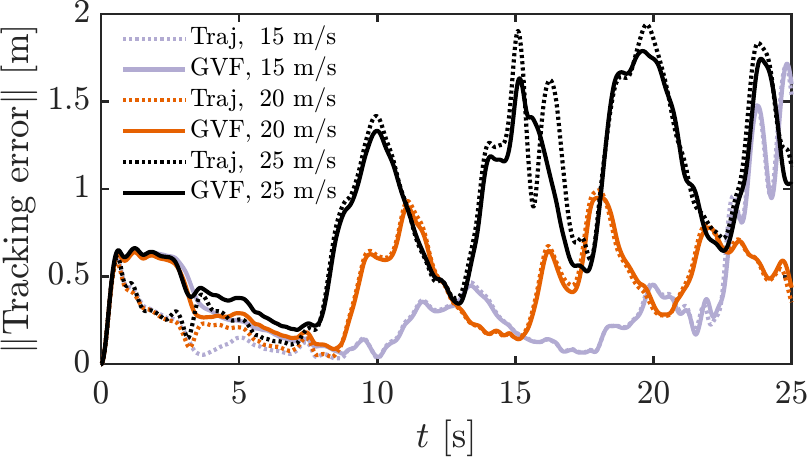}
        \caption{$\| \boldsymbol{v}_{w,s} \| = 10~\mathrm{m/s}$.}
        \label{fig:lissajous_wind}
    \end{subfigure}

    \caption{Tracking error norm on the Lissajous curve. Zero-wind and $\| \boldsymbol{v}_{w,s}\|~=10~\textrm{m/s}$ case.}
    \label{fig:lissajous}
\end{figure*}

\Cref{fig:circle_nowind,fig:lissajous_nowind} compare the tracking performance of the \gls{gvf} guidance and the trajectory guidance for the circular and Lissajous curves, respectively, under various reference speed settings; reference final speed is varied as $V_s \in \{15,20,25\}\,\mathrm{m/s}$. During these maneuvers, acceleration of up to $5g$ is observed. It can be observed that the two methods exhibit comparable performance, with negligible differences. This result is consistent with our expectations.

We also notice that, in contrast to the convergence result proven in \cref{subsec:gvfcontroller}, the tracking error does not converge to zero but instead settles to a small, nonzero value, This is also observed in \cite{tal2023} and can be attributed primarily to two factors. First, the attitude controller \cref{eq:att_controller} does not include a feedforward angular acceleration term and there is input signal inconsistency; $\boldsymbol{q}_c$ is based on the acceleration controller \cref{eq:indi_accel} however $\boldsymbol{\Omega}_\textrm{ff}$ is based purely on the reference flat output signals. Second, the system does not strictly satisfy the double-integrator assumption $\ddot{\boldsymbol{p}} = \boldsymbol{a}_c$ adopted in the guidance controller design, since the rotational dynamics of the vehicle are neglected in this simplification. These assumptions apply to both guidance strategies considered and are common in \gls{uav} control architectures, having a small impact on the overall tracking performance.


\subsubsection{With initial error}
\label{subsubsec:initial_error}
Next, the two guidance strategies are evaluated under a substantial initial position offset from the desired curve. Our hypothesis is that, the \gls{gvf} guidance will outperform the trajectory-tracking approach during the transient. Once convergence is achieved, the behavior is expected to be similar to that described in \cref{subsubsec:no_initial_error}. 

The simulation is conducted on the Lissajous curve, with an initial position offset of $25~\mathrm{m}$ from the curve, along the North axis. The reference speed $s_r$ is chosen with a final value $25~\mathrm{m/s}$, as seen in \cref{fig:sr_25}. For the trajectory guidance, a $10~\mathrm{m/s^2}$ saturation limit is applied to the commanded acceleration due to the tracking error (last term in \cref{eq:ac_traj}) to avoid large acceleration commands at the start of the transient. 

\begin{figure*}[h]
    \centering
    \begin{subfigure}[t]{\columnwidth}
        \centering
        \includegraphics[width=0.85\columnwidth]{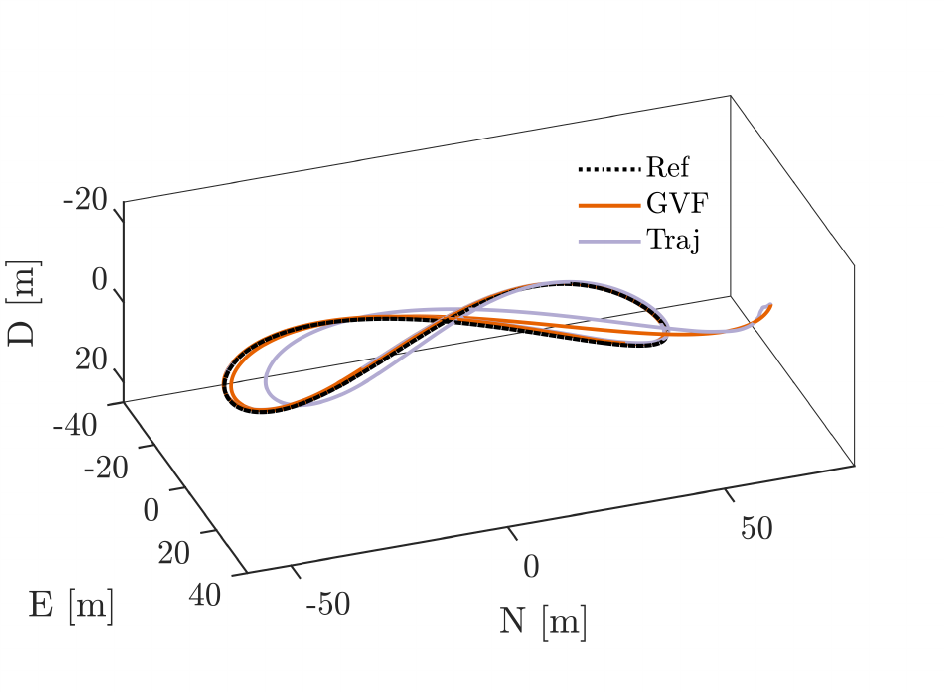}
        \caption{Flight path.}
        \label{fig:initial_groundtrack}
    \end{subfigure}
    \begin{subfigure}[t]{\columnwidth}
        \centering
        \includegraphics[width=0.8\columnwidth]{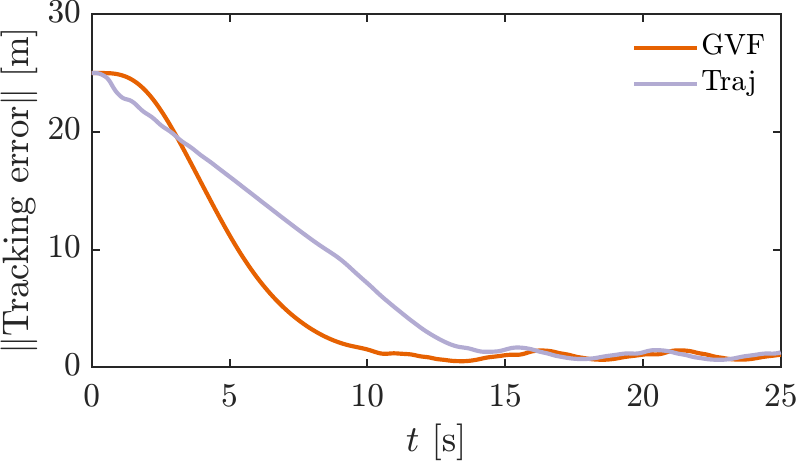}
        \caption{Tracking error norm.}
        \label{fig:initial_error}
    \end{subfigure}
    
    \caption{Tracking performance on the Lissajous curve with initial position error.}
    \label{fig:initial_tracking_perf}
\end{figure*}

\begin{figure*}[h]
    \centering
    \begin{subfigure}[t]{\columnwidth}
        \centering
        \includegraphics[width=0.8\columnwidth]{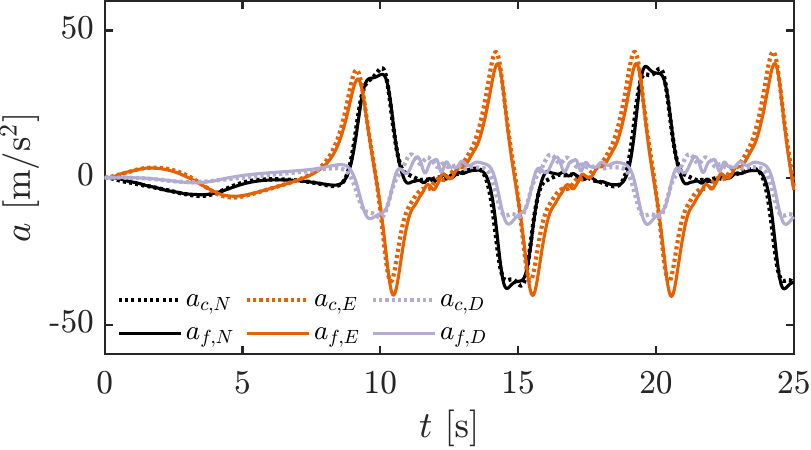}
        \caption{\gls{gvf} guidance.}
        \label{fig:initial_accel_gvf}
    \end{subfigure}
    \begin{subfigure}[t]{\columnwidth}
        \centering
        \includegraphics[width=0.8\columnwidth]{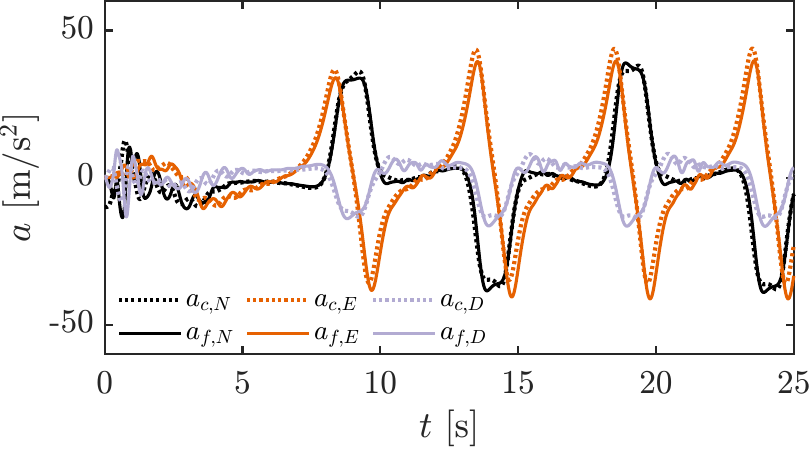}
        \caption{Trajectory guidance.}
        \label{fig:initial_accel_traj}
    \end{subfigure}
    
    \caption{Commanded and actual acceleration on the Lissajous curve with initial position error.}
    \label{fig:initial_accels}
\end{figure*}

\Cref{fig:initial_tracking_perf} illustrates the tracking performance of both guidance strategies in the aforementioned scenario. In the case of the \gls{gvf} guidance, convergence to the curve is noticeably smoother and more controlled. In contrast, the trajectory guidance exhibits an aggressive transient response, with large acceleration commands, as shown in \cref{fig:initial_accels}. This behavior is attributed to the following reasons. First, in the presence of tracking error, the time-dependent feedforward term $\boldsymbol{\Omega}_{\mathrm{ff}}$ becomes inconsistent with the commanded attitude $\boldsymbol{q}_c$, as discussed in \cref{subsubsec:no_initial_error}. As the initial tracking error increases, this inconsistency grows. The second reason is the imposed acceleration saturation. Reducing the saturation limit yields smoother transients at the expense of slower convergence, whereas allowing larger saturation limits may lead to actuator saturation and potential vehicle instability. Once the vehicle converges to the path, the behavior of the two methods is similar. 

This case study highlights an important difference between the two methods. The \gls{gvf} guidance allows for controlled transients, as the predetermined reference speed $s_r$ is applied both during convergence to the path and during traversal. In contrast, with trajectory guidance, a reference speed can only be specified as a feedforward term along the trajectory, but not for the error controller. As a result, an additional saturation mechanism is required, along with its associated limitations, as described above.

\subsection{Performance under unmodelled wind disturbances}
To evaluate the performance of both guidance strategies under wind disturbances, two sources of uncertainty are introduced. 
First, we consider a wind estimator that provides estimates of the mean wind velocity $\hat{\boldsymbol{v}}_{w,s}$. We also consider the presence of an estimation error, such that the estimated mean wind velocity is modeled as
\begin{equation}
    \hat{\boldsymbol{v}}_{w,s}
    =
    R(\varepsilon_{\mathrm{dir}})\,
    (1+\varepsilon_{\mathrm{mag}})
    \boldsymbol{v}_{w,s},
\end{equation}
where $\varepsilon_{\mathrm{mag}} \sim \mathcal{U}(-0.3,\,0.3)$ represents a bounded $\pm30\%$ uncertainty on the wind magnitude, and 
$\varepsilon_{\mathrm{dir}} \sim \mathcal{U}(-0.5,\,0.5)\,\mathrm{rad}$ denotes a bounded estimation error on the wind direction. 
The matrix $R(\varepsilon_{\mathrm{dir}}) \in \mathrm{SO}(2)$ is a planar rotation matrix. Second, the wind gust component $v_{w,g}$ is assumed to be completely unknown to the controller. 

We arbitrarily select a southwest wind. No additional wind directions are explicitly considered, since the simulated maneuvers span all course angles. As a result, the \gls{uav} experiences the wind from all possible relative directions in the body frame. Furthermore, the inclusion of bounded uncertainty in both the wind magnitude and direction ensures that the evaluated scenarios are representative of a wide range of effective wind conditions.

The simulation study is structured as follows. Two wind conditions are considered with wind speeds $\| \boldsymbol{v}_{w,s}\|~\in~\{5,10\}\,\mathrm{m/s}$. For each wind condition, both a low and a high reference final speed are evaluated; $V_s \in \{10,25\}\,\mathrm{m/s}$. It is expected that the \gls{gvf}-based guidance will exhibit superior performance as operating conditions become more demanding, since more aggressive maneuvers induce larger tracking error, which was shown in \cref{subsubsec:initial_error} to negatively affect the trajectory-tracking guidance performance. By more demanding conditions we refer in particular to higher reference speeds, which result in higher-acceleration maneuvers, and stronger wind conditions, which amplify the effect of unmodeled disturbances.

\begin{figure}[h]
    \centering
    \begin{subfigure}[t]{0.49\columnwidth}
        \centering
        \includegraphics[width=\columnwidth]{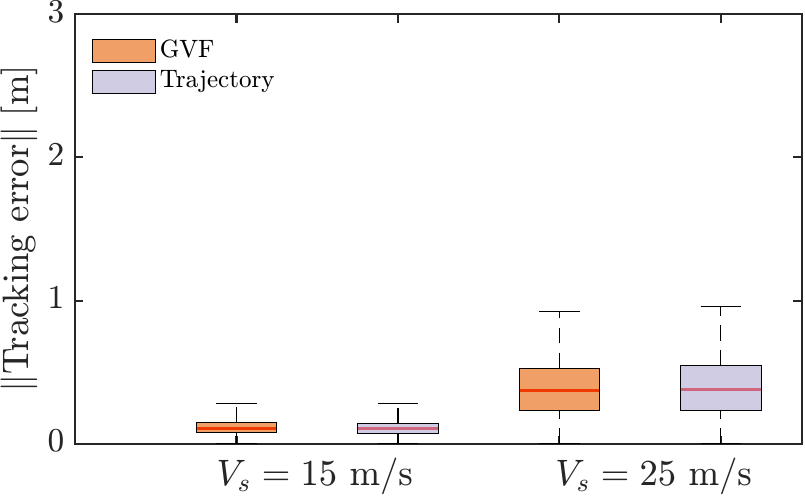}
        \caption{$v_{w,s} = 5~\mathrm{m/s}$.}
        \label{fig:box_circle_lowwind}
    \end{subfigure}\hfill
    \begin{subfigure}[t]{0.49\columnwidth}
        \centering
        \includegraphics[width=\columnwidth]{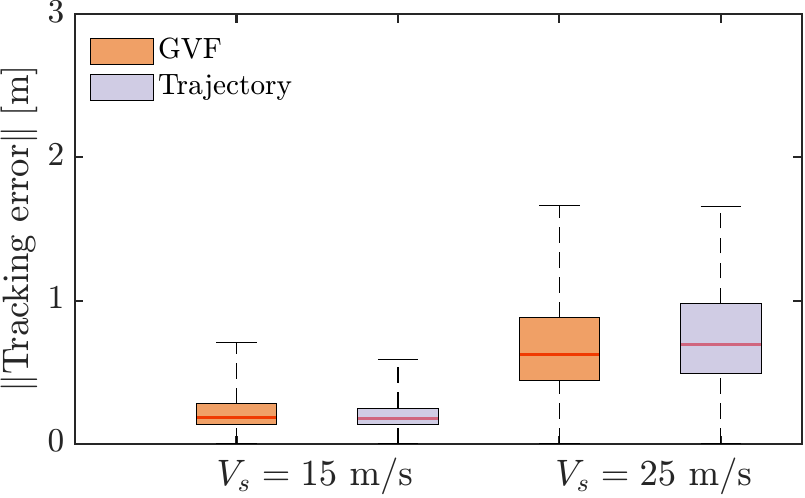}
        \caption{$v_{w,s} = 10~\mathrm{m/s}$.}
        \label{fig:box_circle_highwind}
    \end{subfigure}

    \caption{Statistical analysis of the tracking error norm on the circular curve.} \label{fig:box_circle}
\end{figure}

\begin{figure}[h]
    \centering
    \begin{subfigure}[t]{0.49\columnwidth}
        \centering
        \includegraphics[width=\columnwidth]{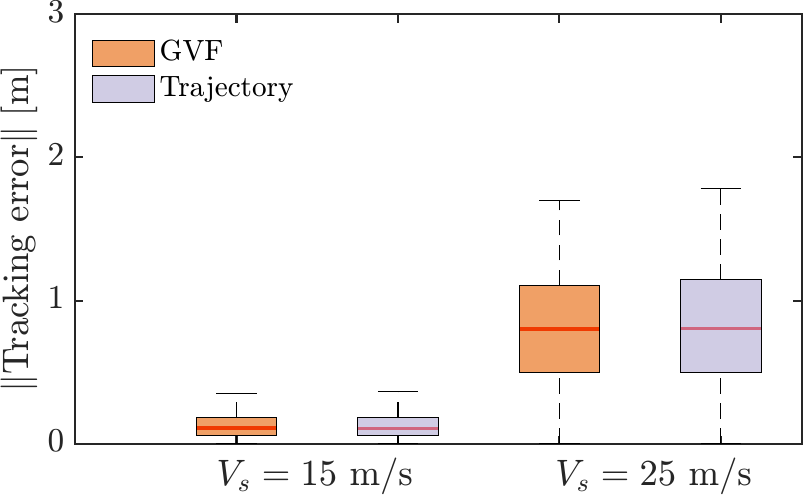}
        \caption{$v_{w,s} = 5~\mathrm{m/s}$.}
        \label{fig:box_lissajous_lowwind}
    \end{subfigure}\hfill
    \begin{subfigure}[t]{0.49\columnwidth}
        \centering
        \includegraphics[width=\columnwidth]{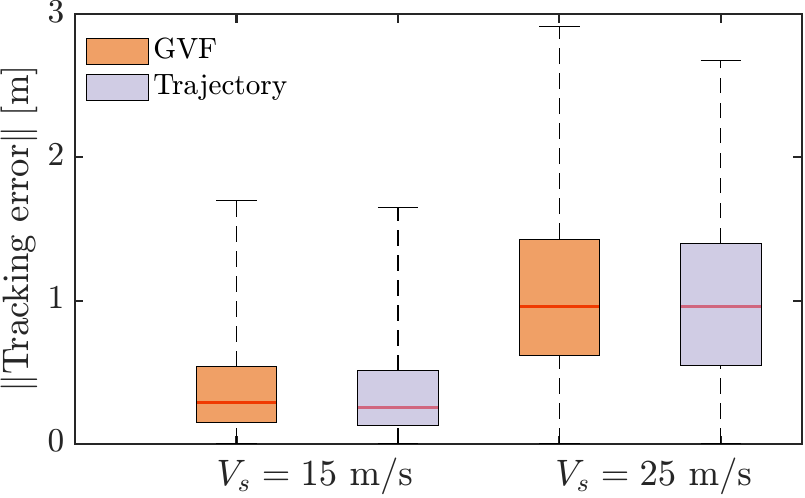}
        \caption{$v_{w,s} = 10~\mathrm{m/s}$.}
        \label{fig:box_lissajous_highwind}
    \end{subfigure}

    \caption{Statistical analysis of the tracking error norm on the Lissajous curve.}
    \label{fig:box_lissajous}
\end{figure}

\Cref{fig:box_circle,fig:box_lissajous} depict the tracking performance of both guidance strategies for the aforementioned scenarios. Increasing the reference speed $s_r$ of the maneuver or the wind speed $\| \boldsymbol{v}_{w,s} \|$ leads to an increase in the median tracking error, and its variability due to the stochastic nature of the wind disturbance, as expected. However, contrary to our initial expectation, the two guidance strategies exhibit very similar performance across all evaluated scenarios, with no statistically or practically significant differences observed. We believe that this result may be attributed to the robust nature of the inner acceleration and attitude \gls{indi} controllers, which can effectively counteract disturbances and unmodelled dynamics.

A representative condition of $\|\boldsymbol{v}_{w,s}\| = 10~\mathrm{m/s}$ is selected, and the norm of the tracking error is illustrated in \cref{fig:circle_wind,fig:lissajous_wind} for various reference final speeds $V_s$ for the circular and Lissajous trajectories, respectively. They are displayed alongside the corresponding nominal zero-wind cases. It is important to note that, in order to achieve the same trajectory in the presence of wind, the motion of the \gls{uav} relative to the airmass differs from that of the zero wind case. As a result, the wind and no-wind cases are not directly comparable. Nevertheless, qualitative conclusions can be drawn. When compared against the nominal zero-wind performance, the normalized tracking error in the presence of wind is consistently larger and exhibits increased variability. Furthermore, in agreement with the results shown in \cref{fig:box_circle,fig:box_lissajous}, both guidance strategies yield very similar normalized tracking errors.


\section{CONCLUSION}
In this work we developed a guidance controller based on a \gls{gvf} and formally integrated it with a state-of-the-art acceleration and attitude control framework for tailsitter \glspl{uav}. By modifying the parametric formulation of the \gls{gvf}, we proved exponential stability of the tracking error dynamics for a single integrator system, thereby enabling consistent tuning and comparison between the \gls{gvf}-based guidance and a traditional trajectory-tracking guidance controller.

By simulating a realistic tailsitter model under uncertainty induced by unmodelled wind effects, we observed that, for agile flight maneuvers in which the \gls{uav} is expected to initiate the motion with a small position error and where no significant deviation from the path is anticipated due to sources other than external wind disturbances, the two guidance strategies exhibit very similar performance, with no practically significant advantage of either method. In such scenarios, the added complexity of the \gls{gvf} guidance may not be justified.

However, the \gls{gvf}-based guidance exhibits a clear advantage in the presence of substantial deviations from the desired path, as may occur in practical \gls{uav} applications. In such cases, in contrast to traditional trajectory-tracking guidance, the system exhibits increased robustness, characterized by smooth and well-behaved convergence toward the desired path while avoiding aggressive transients and large commanded accelerations.

While existing \gls{gvf}-based approaches primarily target fixed-wing aircraft, this work demonstrates that the proposed guidance strategy is also applicable to hybrid \glspl{uav} and rotorcraft platforms. The developed \gls{gvf} guidance is currently applicable to continuous maneuvers with $s_r \neq 0$. As future work, we propose extending the formulation to enable point-to-point motion with stopping capability, thereby increasing the generality of the method.

Finally, this work extends the differential flatness transform for a tailsitter to explicitly incorporate wind velocity knowlwdge, enabling active wind compensation. A next step could include the development of a wind estimator based on the $\phi$-theory model, as well as the adoption of a sideslip-based control formulation in place of yaw-angle control. These extensions can enable more robust operation of tailsitter platforms in windy environments.

\section*{ACKNOWLEDGMENTS}
The authors would like to thank H. de Marina and J. Bautista for their assistance and the discussions.

\appendix

In this appendix, we provide analytical expressions for the matrices introduced in \cref{subsec:gvfcontroller} to facilitate a better understanding of the developed method. The parametric path and its derivative w.r.t. $w$ are given by
$\boldsymbol{f}=\begin{bmatrix} f_1 & f_2 & f_3 \end{bmatrix}^\top$
and
$\boldsymbol{f'}=\begin{bmatrix} f_1' & f_2' & f_3' \end{bmatrix}^\top$,
respectively. Similarly, the path tracking error is $\boldsymbol{\phi}=\begin{bmatrix}
        \phi_1 & \phi_2 & \phi_3
    \end{bmatrix}^\top$ and the \gls{gvf} gain $K=\mathrm{diag}(k_1, k_2, k_3)$.
Then the \gls{gvf} is explicitly expressed by:
\begin{equation}
    \boldsymbol{\chi} =\begin{bmatrix}
        \boldsymbol{\chi}_p \\ \chi_w
    \end{bmatrix}
    =\begin{bmatrix}
        -f'_1 + f'_1\Sigma - k_1\phi_1 \\
        -f'_2 + f'_2\Sigma - k_2\phi_2 \\
        -f'_3 + f'_3\Sigma - k_3\phi_3 \\
        -1 + \Sigma
    \end{bmatrix},
\end{equation}
where 
\begin{equation}
    \begin{aligned}
        \Sigma = (K\boldsymbol{\phi})^\top\boldsymbol{f}' = k_1\phi_1f_1' + k_2\phi_2f_2' + k_3\phi_3f_3'.
    \end{aligned}
\end{equation}
Analytical expressions of the matrices used in the computation of $\dot{\boldsymbol{\chi}}_p$ and $\dot{\chi}_w$ are given below:
% \begin{equation}
% \mathbf{J}_{\chi,\xi} =
%     \begin{bmatrix}
%         \mathbf{J}_{\chi_p,p} & \frac{\partial \boldsymbol{\chi}_{p}}{\partial w}\\
%         \mathbf{J}_{\chi_w,p} & \frac{\partial \chi_{w}}{\partial w}
%     \end{bmatrix},
% \end{equation}
\begin{equation}
    \mathbf{J}_{\chi_p,p} = 
    \begin{bmatrix}
    -k_1+f_1'^2k_1 & f_1'f_2'k_2 & f_1'f_3'k_3  \\
    f_2'f_1'k_1 & -k_2+f_2'^2k_2 & f_2'f_3'k_3  \\
    f_3'f_1'k_1 & f_3'f_2'k_2 & -k_3+f_3'^2k_3 
    \end{bmatrix}
\end{equation}
\begin{equation}
    \mathbf{J}_{\chi_w,p} = 
    \begin{bmatrix}
    k_1f_1' & k_2f_2' & k_3f_3'
    \end{bmatrix}
\end{equation}
\begin{equation}
    \frac{\partial \boldsymbol{\chi}_p}{\partial w} = 
    \begin{bmatrix}
    -f_1''+k_1f_1'+f_1''\Sigma + f_1'\Sigma' \\
    -f_2''+k_2f_2'+f_2''\Sigma + f_2'\Sigma' \\
    -f_3''+k_3f_3'+f_3''\Sigma + f_3'\Sigma'
    \end{bmatrix}
\end{equation}
\begin{equation}
    \frac{\partial \chi_w}{\partial w} = 
        \Sigma',
\end{equation}
where
\begin{equation}
    \begin{aligned}
        \Sigma' &= k_1\phi_1f_1'' - k_1f_1'+ k_2\phi_2f_2'' - k_2f_2' \\
 &\quad+ k_3\phi_3f_3'' - k_3f_3'.
    \end{aligned}
\end{equation}
Next we present the analytical expressions of the matrices used in the computation of $\ddot{\boldsymbol{\chi}}_p$:
\begin{equation}
    \frac{\partial^2\boldsymbol{\chi}_p}{\partial w^2} = 
\begin{bmatrix}
     -f_1'''+k_1f_1''+f_1'''\Sigma + 2f_1''\Sigma' + f_1'\Sigma'' \\
     -f_2'''+k_2f_2''+f_2'''\Sigma + 2f_2''\Sigma' + f_2'\Sigma'' \\
     -f_3'''+k_3f_3''+f_3'''\Sigma + 2f_3''\Sigma' + f_3'\Sigma''
\end{bmatrix},
\end{equation}
where
\begin{equation}
    \begin{aligned}
        \Sigma'' &= k_1\phi_1f_1''' - 3k_1f_1'f_1'' + k_2\phi_2f_2''' - 3k_2f_2'f_2'' \\ &\quad +k_3\phi_3f_3''' - 3k_3f_3'f_3''.
    \end{aligned}
\end{equation}
\begin{equation}
\begin{aligned}
    \mathbf{H}_{\chi_p,pw} = \quad\quad\quad\quad\quad\quad\quad\quad\quad\quad\quad\quad\quad\quad\quad\quad\quad\quad\quad\quad\\
    \begin{bmatrix}
        2k_1f_1'f_1'' & k_2(f_1''f_2'+f_1'f_2'') & k_3(f_1''f_3'+f_1'f_3'')\\
         k_1(f_2''f_1'+f_2'f_1'') &  2k_2f_2'f_2'' & k_3(f_2''f_3'+f_2'f_3'')\\
         k_1(f_3''f_1'+f_3'f_1'') & k_2(f_3''f_2'+f_3'f_2'')  & 2k_3f_3'f_3''
    \end{bmatrix}.
\end{aligned}
\end{equation}
For the calculation of $\dot{\boldsymbol{\chi}}_{p_g}$ the analytical expressions presented below are used:
\begin{equation}
    \mathbf{J}_{\chi_{p_g},p} = 
    \begin{bmatrix}
    f_1'^2k_1 & f_1'f_2'k_2 & f_1'f_3'k_3  \\
    f_2'f_1'k_1 & f_2'^2k_2 & f_2'f_3'k_3  \\
    f_3'f_1'k_1 & f_3'f_2'k_2 & f_3'^2k_3
    \end{bmatrix}
\end{equation}
\begin{equation}
    \frac{\partial \boldsymbol{\chi}_{p_g}}{\partial w} = 
    \begin{bmatrix}
    -f_1''+f_1''\Sigma' + f_1'\Sigma' \\
    -f_2''+f_2''\Sigma' + f_2'\Sigma' \\
    -f_3''+f_3''\Sigma' + f_3'\Sigma'
    \end{bmatrix}.
\end{equation}
Finally, we provide the expressions of the matrices involved in the calculation of $\ddot{\boldsymbol{\chi}}_{p_g}$:
\begin{equation}
    \frac{\partial^2\boldsymbol{\chi}_{p_g}}{\partial w^2} = 
\begin{bmatrix}
     -f_1'''+f_1'''\Sigma + 2f_1''\Sigma' + f_1'\Sigma'' \\
     -f_2'''+f_2'''\Sigma + 2f_2''\Sigma' + f_2'\Sigma'' \\
     -f_3'''+f_3'''\Sigma + 2f_3''\Sigma' + f_3'\Sigma''
\end{bmatrix}
\end{equation}
\begin{equation}
\begin{aligned}
    \mathbf{H}_{\chi_{p_g},pw} = \mathbf{H}_{\chi_{p},pw}. 
\end{aligned}
\end{equation}

% \textcolor{orange}{To be removed}:
% \begin{equation}
% \begin{aligned}
%     \ddot{\boldsymbol{\chi}}_{p} &= \mathbf{H}_{\chi_{p},pp}[\dot{\boldsymbol{p}},\dot{\boldsymbol{p}}] + \mathbf{H}_{\chi_{p},wp}[\dot{w},\dot{\boldsymbol{p}}] + \mathbf{J}_{\chi_{p},p}\ddot{\boldsymbol{p}} \\
%     &+\mathbf{H}_{\chi_{p},pw}[\dot{\boldsymbol{p}},\dot{w}] + \mathbf{H}_{\chi_{p},ww}[\dot{w},\dot{w}] + \mathbf{J}_{\chi_{p},w}\ddot{w}.
% \end{aligned}
% \end{equation}
% $$\psi = \chi - \delta_w$$,
% where $\chi$ is the course angle and $\delta_w$ the wind correction angle

\addtolength{\textheight}{-12cm}  % This command serves to balance the column lengths
                                  % on the last page of the document manually. It shortens
                                  % the textheight of the last page by a suitable amount.
                                  % This command does not take effect until the next page
                                  % so it should come on the page before the last. Make
                                  % sure that you do not shorten the textheight too much.

\end{document}